\documentclass[reprint,aps,pra,superscriptaddress,showpacs,nofootinbib]{revtex4-2}
\usepackage{blindtext}
\usepackage{graphicx}% Include figure files
\usepackage{dcolumn}% Align table columns on decimal point
\usepackage{bm}% bold math
\usepackage{color}
\usepackage{ulem}
\usepackage{soul}
\usepackage{siunitx}
\usepackage{amsmath}
\usepackage[english]{babel}
\usepackage[symbol]{footmisc}

\begin{document}
\title{Interferometric measurement of the deflection of light by light in air}

\author{Adrien E. Kraych}
\email{adrien.kraych@ijclab.in2p3.fr}
\author{Aur\'elie Max Mailliet}
\author{Fran\c{c}ois Couchot}
\author{Xavier Sarazin}
\author{Elsa Baynard}
\author{Julien~Demailly}
\author{Moana Pittman}
\affiliation{Université Paris-Saclay, CNRS/IN2P3, IJCLab, 91405 Orsay, France}
\author{Arache Djannati-Ata\"{i}}
\affiliation{Universit\'e Paris Diderot, CNRS/IN2P3, APC, Paris, France}
\author{Sophie Kazamias}
\affiliation{Université Paris-Saclay, CNRS/IN2P3, IJCLab, 91405 Orsay, France}
\author{Scott Robertson}
\affiliation{Institut Pprime, CNRS -- Université de Poitiers -- ISAE-ENSMA, 86073 Poitiers, France}
\author{Marcel Urban}
\affiliation{Université Paris-Saclay, CNRS/IN2P3, IJCLab, 91405 Orsay, France}

\begin{abstract}
The aim of the DeLLight (Deflection of Light by Light) experiment is to observe for the first time the optical nonlinearity in vacuum, as predicted by Quantum Electrodynamics, by measuring the refraction of a low-intensity focused laser pulse (probe) when crossing the effective vacuum index gradient induced by a high-intensity focused laser pulse (pump). 
The deflection signal is amplified by using a Sagnac interferometer.
Here, we report the first measurement performed with the DeLLight pilot interferometer, of the deflection of light by light in air, with a low-intensity pump. We show that the deflection signal measured by the interferometer is amplified, and is in agreement with the expected signal induced by the optical Kerr effect in air. Moreover, we verify that the signal varies as expected as a function of the pump intensity, the temporal  delay between the pump and the probe, and their relative polarisation. These results represent a proof of concept of the DeLLight experimental method based on interferometric amplification. 

\end{abstract}

\maketitle

\section{Introduction}

In the vacuum of classical electrodynamics, the linearity of Maxwell’s equations 
forbids any self-interaction of the electromagnetic field. The permittivity and permeability of free space -- and, consequently, the speed of light -- do not depend on the presence of other electromagnetic fields. However, in Quantum Electrodynamics (QED), vacuum is filled with fluctuations of both photons and electron-positron pairs, with the latter inducing a weak effective interaction between real photons. The QED vacuum is thus expected to behave as a non-linear optical medium when it is stressed by intense electromagnetic fields, as predicted initially by Euler and Heisenberg~\cite{euler1935streuung,heisenberg1936folgerungen} and formulated later within the QED framework by Schwinger~\cite{schwinger1951gauge} as photon-photon scattering (four-wave interactions). 

%Although 
Different occurrences of the photon-photon scattering process have already been observed in the sense of particle scattering (at the Stanford Linear Accelerator Center~\cite{burke1997positron} and at the Large Hadron Collider~\cite{atlas2017evidence,sirunyan2019evidence}) and it is still the subject of intense research involving the scattering of electron beams with intense laser pulses~\cite{heinemann2020luxe,abramowicz2021conceptual,salgado2021single,fedotov2023advances}.
However, no experiment has yet been able to demonstrate the nonlinear optical signature of the vacuum on a macroscopic scale, i.e., a coherent phenomenon corresponding to a pure undulatory process at large scale and treated classically in the long-wavelength limit.
The optical nonlinearity of the vacuum gives rise to a number of new optical effects, still to be observed: vacuum birefringence \cite{klein1964birefringence,baier1967vacuum,homma2011probing,heinzl2006observation}, harmonic generation in vacuum \cite{di2005harmonic,lundin2006analysis}, interference effects \cite{king2010double,tommasini2010light} (see also references in~\cite{di2012extremely,narozhny2015extreme,king2016measuring}). 
Up to now, the main experimental efforts have involved testing vacuum magnetic birefringence in the presence of an external magnetic field~\cite{ejlli2020pvlas,cadene2014vacuum,fan2017oval,chen2007q}. In particular, the PVLAS experiment, using a $2.5$~T magnetic field, has achieved the best sensitivity so far, reaching an experimental uncertainty about one order of magnitude above the predicted QED effect after about 100~days of collected data~\cite{ejlli2020pvlas}.

In order to observe for the first time the optical nonlinearity of the vacuum, the DeLLight collaboration aims to directly observe a change of the vacuum refractive index using strong electromagnetic fields contained in high-intensity ultra-short laser pulses delivered by the LASERIX facility (1.5~J per pulse, each of $40$~fs duration, with a $10$~Hz repetition rate). This % optical 
phenomenon is similar to the optical Kerr effect in an optical material medium, corresponding to a variation $\delta n$ of the refraction index of the medium, induced by the intense field of the laser pulse passing through the medium, where $\delta n$ is at first order proportional to the field intensity $I$~\cite{boyd2008nonlinear}. 

The principle of the DeLLight experiment, as reported in~\cite{sarazin2016refraction,robertson2021experiment}, is to measure the refraction of a low-intensity focused laser pulse when crossing the vacuum optical index gradient produced by an intense focused counter-propagating laser pulse. 
As the expected deflection angle is challengingly small, the deflection signal is amplified by using a Sagnac interferometric measurement. 
The amplification is based on the so called weak value amplification method, proposed by Aharanov {\it et al.} in 1988~\cite{aharonov1988result,aharonov2010time}, and more recently developed to measure sub-picoradian rotation of a mirror in a Sagnac interferometer using a continuous laser beam in the search for gravitational anomalies at short distance~\cite{dixon2009ultrasensitive,egan2012weak}.

Before starting the DeLLight measurements in vacuum with intense laser pulses, 
a pilot experiment has been developed, running in air with relatively low-intensity laser pulses. The goal is to demonstrate the feasibility of the DeLLight experimental method, by measuring, via a Sagnac interferometer, the refraction of a probe pulse crossing the optical Kerr index gradient induced in air by a low-energy pump pulse. In particular, the current work aims to verify the principle of interferometric amplification in the framework of the DeLLight experiment by measuring the well-known optical Kerr effect in air.
In this article, we report the results of these measurements. We show that the deflection signal is well amplified, thanks to the interferometric measurement, and in agreement with the expected amplification factor. 
We also measure the amplified deflection signal as a function of the pump energy, the temporal delay between the pump and the probe, and of their relative polarisation. 

\section{Description of the Dellight pilot experiment}

\begin{figure*}
    \centering
    \includegraphics[width=18cm]{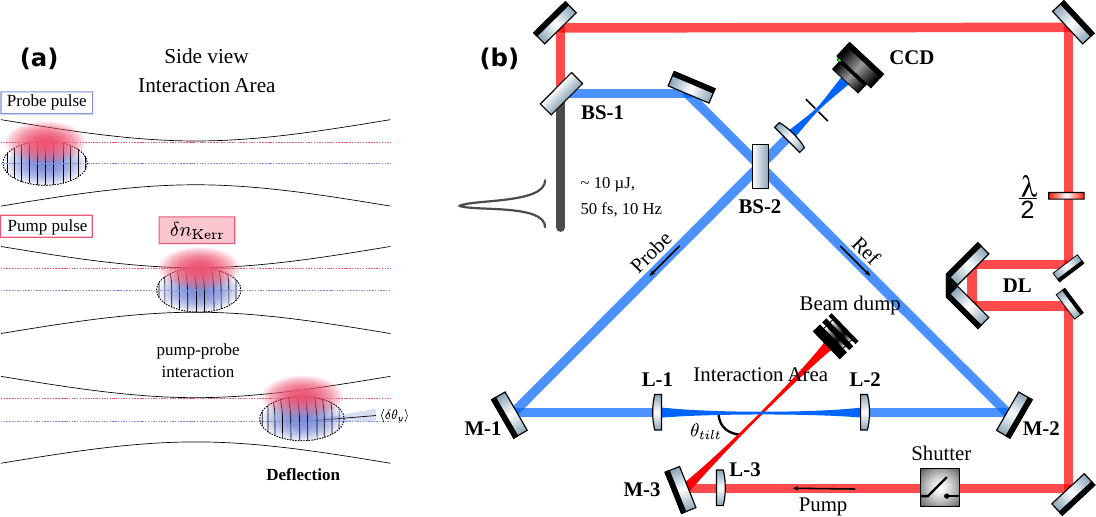}
    \caption{(a) Schematic view of the interaction between the probe pulse (in blue) and the co-propagating pump pulse (in red), seen from the side. The reference pulse is not represented in this figure as it is not in time coincidence at the interaction point and thus remains unaffected by both the pump and the probe. The lines inside the probe pulse correspond to the wave fronts, which are gradually rotated by the Kerr index gradient induced by the pump. (b) Schematic view of the DeLLight pilot experiment setup.   }
    \label{fig:setup}
\end{figure*}

\subsection{Interaction Area}

In vacuum, the nonlinear QED modification of the vacuum optical index is maximum when the pump and the probe pulses are propagating in opposite directions, as shown for instance in~\cite{robertson2019optical}.
On the other hand, in an optical material medium like air, 
the nonlinear optical Kerr effect generated by the contribution of two synchronized laser pulses is locally independent from their relative direction of propagation. Therefore, in order to maximise the interaction length $L_{int}$ of the probe and the pump pulses and thus maximize the deflection of the probe pulse in air, the pump and the probe are here co-propagating. 
The deflection of the probe induced by the pump is then integrated along the interaction length, until the two pulses are spatially separated due to the tilt angle $\theta_{\rm tilt}$ between the pump and the probe beam. 

As shown in Figure~\ref{fig:setup}(a), the probe pulse propagating simultaneously with the pump will react to the refractive index $\delta n_{\rm Kerr} = n_2 \times I_{\rm pump}$ induced by the pump, where $ I_{\rm pump}$ is its intensity in W/cm$^2$ (integrated over the pulse envelop), and $n_2$ is the Kerr index of air. The pump beam is shifted vertically with respect to that of the probe beam by a distance $b$, named impact parameter, such that the refractive index profile generated by the pump is transversely asymmetric as seen by the probe. This will result in an average deflection of the probe pulse through an angle $\langle \theta_y \rangle$ in the direction of increasing optical index, i.e. towards the pump beam axis. As shown in~\cite{robertson2021experiment}, for Gaussian pulses, the deflection angle is maximum when 
$b=b_{\rm opt}=\sqrt{w_0^2+W_0^2}/2$ 
where $w_0$ and $W_0$ are the waists at focus of the probe and pump, respectively. 

\subsection{The Sagnac interferometer}

The setup of the DeLLight pilot experiment, shown in Figure~\ref{fig:setup}(b), uses a Sagnac interferometer in order to amplify the vertical shift of the deflected probe pulse.

The initial laser pulse used for the pilot experiment is first filtered by a spatial filter composed of two lenses and a pinhole at focus in order to obtain a transverse intensity profile close to a gaussian profile. The pulse duration is $\Delta t \simeq 70$~fs, the central wavelength is $\lambda=810$~nm with a bandwidth $\Delta \lambda \simeq 40$~nm, the  transverse diameter at half-maximum (FWHM) is about 1~mm, the repetition rate is $10$ Hz and the energy $E$ is at most $50 \text{~} \mu$J. This pulse is split upstream by a beamsplitter (BS-1). 
The transmitted pulse is used as the pump pulse.
The reflected pulse, referred to as the incident pulse, is sent into the Sagnac interferometer with an energy of 2~µJ per pulse and an intensity $I_{in}$. It is split at the input of the interferometer by a 50/50 beamsplitter (Semrock FS01-BSTiS-5050P-25.5; BS-2 on Figure~\ref{fig:setup}), which generates two pulses (the probe and the reference) propagating into the Sagnac interferometer along the same path in opposite directions.

The Sagnac interferometer is in a right-angled isosceles triangle configuration, formed by the beamsplitter and two dielectric mirrors (M-1 and M-2). Both counter-propagating pulses are focused in the interaction area, then recollimated via two optical lenses (L-1 and L-2) of equal focal length $f=100$~mm separated by the distance $d=2f$. The lenses are placed between the two mirrors. The pump pulse is also focused in the interaction area by a separate optical lens (L-3) of the same focal length $f$.

In this design, the probe pulse co-propagates with the pump.
A delay stage (DL in Figure~\ref{fig:setup}) ensures the time coincidence of the probe and the pump pulses in the interaction area.  The reference pulse (Ref) is not in time coincidence at the interaction point and is therefore unaffected by the pump. 

In the absence of the pump and with a perfectly aligned Sagnac interferometer, the two counter-propagating probe and reference pulses are 
in opposite phase in the dark output of the interferometer, and interfere destructively. 
However, the beamsplitter of the interferometer is not exactly symmetric in reflection and transmission, and the phase noise is never totally null. Therefore the extinction of the interferometer is limited and the residual interference intensity profile $I_{out}$ in the dark output is measured by a CCD camera (Basler acA3088-16gm, pixel size $5.84 \times 5.84$ $\mu$m$^2$).
The extinction is quantified by the extinction factor $\mathcal{F}$, defined as $\mathcal{F} = I_{out}/I_{in}$. 
A spatial filter is placed in the dark output in front of the CCD in order to suppress the high spatial frequency components of the signal, induced by the diffusion on the surface defects of the optics inside the interferometer being responsible for a large amount of the phase noise.
The filter is composed of an optical lens (focal length $f = 200$ mm) and a pinhole of diameter $150$ µm placed at the focus of the lens, corresponding to an angular cutoff of 375~$\mu$rad. 

When the pump pulse interacts with the probe pulse, the wavefront of the later is refracted by the induced Kerr index gradient $\delta n$, while the reference pulse is unaffected. 
After recollimation by the second lens, the  refracted probe pulse is then transversally vertically shifted with respect to the unrefracted reference pulse by an average distance $\langle \delta y \rangle = \langle \theta_y \rangle f$, where $\langle \theta_y \rangle$ is the average deflection angle of the refracted probe pulse due to $\delta n$. 
The probe pulse is also phase shifted by an average phase delay $\delta \psi = 2\pi \delta n L_{int}/\lambda$, with respect to the reference pulse.
The interference of the refracted probe pulse with the reference pulse in the dark output produces a transverse vertical displacement $\Delta y$ of the interference intensity profile, which is measured by the CCD camera.
As explained in the following part, the advantage of the interferometric measurement relies on the amplification of the displacement signal $\Delta y$ as compared to the direct signal $\delta y$ which would be measured by using a standard pointing method. 
This is the guiding principle behind the use of the Sagnac interferometer.

\subsection{Amplified signal in the dark output}
\label{sec:amplification}

In the absence of pump interaction (``OFF'' measurement), it is shown in Appendix~\ref{ap:analytical-calculation} that the intensity profile $I_{\rm OFF}$ in the dark output of the Sagnac interferometer is given by:
\begin{equation}
\label{eqIOFFgen}
    I_{\rm OFF}(x, y) = \left( (\delta a)^2 + (\delta \phi(x,y))^2 \right) I_{\rm in}(x,y)
\end{equation}
The parameter $\delta a$ characterizes the asymmetry between the reflection $R$ and transmission $T$ coefficients in intensity of the beamsplitter, with $R=(1-\delta a)/2$ and $T=(1+\delta a)/2$. 
The parameter $\delta \phi(x,y)$ is the phase noise between the probe and the reference in the dark output of the Sagnac interferometer. It is either related to an intrinsic asymetry of the  beamsplitter or a phase noise induced by the surface defects of the optics inside the interferometer with a transversal dependence $(x, y)$. 
The extinction factor is then equal to $\mathcal{F}=(\delta a)^2 + (\delta \phi(x,y))^2$.

When the pump interacts with the probe (``ON'' measurement), it is shown in Appendix~\ref{ap:analytical-calculation} that the intensity profile $I_{\rm ON}$ in the dark output becomes (when $\delta a \ll 1$):
\begin{align}
\begin{split}\label{eqIONgen}
    I_{\rm ON}(x, y) ={}&  (\delta a)^2 I_{\rm in} \left(x,y + \frac{1}{2\delta a} \delta y \right)\\
                    & + \left(\delta \phi(x,y) + \frac{\delta \psi}{2} \right)^2  I_{\rm in}\left(x,y-\frac{\delta y}{2}\right)
\end{split}
\end{align}
The first term of Equation~(\ref{eqIONgen}) is the deflection signal and corresponds to the amplified displacement of the interference profile barycenter by a distance $\delta y/(2\delta a)$. The second term is related to the phase delay $\delta \psi$ induced by the pump, which produces an intensity variation of the interference signal. 

If the contribution of the phase term is negligible ($(\delta \phi+\delta\psi/2)^2 \ll (\delta a)^2$), the intensity profiles become:
\begin{eqnarray}
\label{eqION_deflex}
    I_{\rm ON}(x, y) & = & (\delta a)^2 I_{\rm in} \left(x,y + \frac{\delta y}{2\delta a}\right) \\
    I_{\rm OFF}(x, y) & = & (\delta a)^2 I_{\rm in} (x,y) \\
    \mathcal{F} & = & (\delta a)^2
\end{eqnarray}
Hence, the probe is deflected and transversally shifted by a distance $\delta y$, the interference intensity profile barycenter in the dark output is vertically shifted by a distance $\Delta y = - \delta y /(2\delta a)$. 
The signal is therefore amplified by an amplification factor $\mathcal{A}$ given by: 
\begin{eqnarray}
\label{facteurampli}
\mathcal{A} = \frac{\Delta y}{\delta y} = -\frac{1}{2\delta a} = - \frac{1}{2\sqrt{\mathcal{F}}}
\end{eqnarray}
The amplification factor is large when $\delta a \ll 1$,  corresponding to a high reflection/transmission symmetry of the beamsplitter and a strong extinction of the interferometer.

When the contribution of the phase noise is not negligible, the second term in Equation~(\ref{eqIONgen}) can shift the barycenter of the intensity profile. Indeed, the measured displacement signal $\Delta y$ is no longer solely due to the deflection (i.e., equal to $\delta y /(2\delta a)$) but becomes biased by the phase noise transverse distribution, and the amplification factor is reduced.

\subsection{Extinction factor}

\begin{figure}
    \centering
    \includegraphics[width=8.5cm]{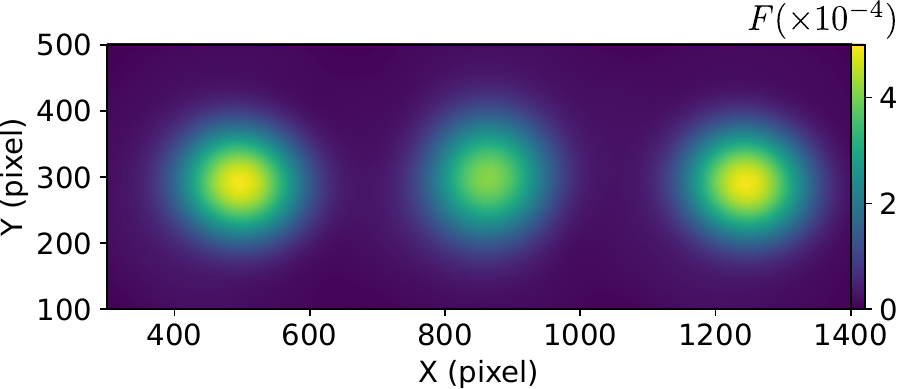}
    \caption{ Intensity profile recorded by the CCD camera in the dark output of the interferometer. The observed
intensity has been normalized by the maximum intensity of the input pulse such that the observed intensity corresponds to the extinction factor.}
    \label{fig:darport}
\end{figure}

A typical transverse intensity profile recorded by the CCD camera in the dark output of the interferometer, after alignment of the interferometer, is shown in Fig.~\ref{fig:darport}. 
The central spot corresponds to the interference signal. 
The two other spots observed on opposite lateral sides are due to back-reflections on the rear side of the beamsplitter. Their respective intensities are $I_{AR,1} = I_{AR,2} = (R_{AR}/2)  I_{in}$,  where $R_{AR} = 10^{-3}$ is the back-reflection coefficient of the beamsplitter.
As explained in Appendix~\ref{ap:back-reflections}, one back-reflection ($I_{AR,1}$) corresponds to the direct back-reflection of the probe pulse. This beam is therefore deflected by the pump interaction and is used to measure the direct deflection signal $\delta y$. The second back-reflection ($I_{AR,2}$) is not deflected by the pump and corresponds to the direct image of the incident beam. It is used to measure and suppress the beam pointing fluctuations, as detailed below.

With the current available beamsplitter, the extinction factor is limited by the presence of the back-reflections. The extinction is tuned in order to obtain an interference intensity of the same order of magnitude as the back-reflection intensities. 
It is done by rotating the beamsplitter of the interferometer by 1 degree in the horizontal plane changing the incident angle of the laser pulse from 45 to 46~degrees. 
At this incident angle, the measured transmission and reflection coefficients are $R=49 \%$ and $T=51 \%$, corresponding to the asymmetry coefficient $\delta a = 0.02$, an extinction factor $\mathcal{F} =(\delta a)^2 = 4 \times 10^{-4}$ and an expected negative amplification factor $\mathcal{A_F} = -25$ in the case of a negligible phase noise. Here, $\mathcal{A_F}$ is defined by the extinction factor such that $\mathcal{A_F} = - \frac{1}{2\sqrt{\mathcal{F}}}$ unlike $\mathcal{A}$ which is defined as $\mathcal{A} = \frac{\Delta y}{\delta y}$. For a negligible phase noise, we have $\mathcal{A_F} = \mathcal{A}$.

\subsection{Signal analysis method}

The analysis method to measure the deflection signal has been described in~\cite{robertson2021experiment}. We summarize here the method. 
The deflection signal is measured by alternating laser shots with and without interactions between the pump and the probe pulse (ON and OFF measurements).
This is done by the use of a fast iris motorized shutter which is synchronized at 5~Hz with the laser shots. 
For each measurement, the barycenters $\bar{y}_{\mathrm{sig}}$ and $\bar{y}_{\mathrm{ref}}$ of the intensity profiles of the interference signal $I_{out}$ and the back-reflection $I_{AR,2}$ respectively,  are calculated along the vertical axis, using a square analysis window (or {\it Region of Interest}, RoI), whose size is equal to half the width (FWHM) of the transverse intensity profile. 

The beam pointing fluctuations are suppressed for each ON and OFF measurement $i$, by using the correlation of $\bar{y}_{\mathrm{sig}}$ and $\bar{y}_{\mathrm{ref}}$.
One obtains the corrected positions:
\begin{eqnarray}
\bar{y}_{\mathrm{corr}}^{\mathrm{OFF}}(i) & = & \bar{y}_{\mathrm{sig}}^{\mathrm{OFF}}(i) - \left( a^{\mathrm{OFF}} \bar{y}_{\mathrm{ref}}^{\mathrm{OFF}}(i) + b^{\mathrm{OFF}} \right) \nonumber \\
\bar{y}_{\mathrm{corr}}^{\mathrm{ON}}(i) & = & \bar{y}_{\mathrm{sig}}^{\mathrm{ON}}(i) - \left( a^{\mathrm{OFF}}  \bar{y}_{\mathrm{ref}}^{\mathrm{ON}}(i) + b^{\mathrm{OFF}} \right) 
\label{eq:BPcorrections}
\end{eqnarray}
where $a^{\mathrm{OFF}}$ and $b^{\mathrm{OFF}}$ are obtained by fitting the linear correlation, using only the OFF measurements. 
The amplified signal $\Delta y (i)$ of the ``ON-OFF'' measurement $i$ is then given by
\begin{eqnarray}
\Delta y (i) = \bar{y}_{\mathrm{corr}}^{\mathrm{ON}}(i) - \bar{y}_{\mathrm{corr}}^{\mathrm{OFF}}(i).
\end{eqnarray}
The amplified deflected signal $\langle \Delta y\rangle$ is obtained in the presented results by averaging 200 successive ON-OFF measurements $\Delta y (i)$, corresponding to 400 successive laser shots and a 40~seconds measurement duration (10 Hz repetition rate).

The direct deflection signal $\langle \delta y \rangle$ is measured following the same procedure but using the barycenter of the back reflection intensity profile $I_{AR,1}$ instead of the barycenter of the interference intensity profile.

\section{Results}

We have measured the deflection as a function of four different parameters: the delay between the arrival at focus of the pump and probe pulses, the pump intensity, the relative polarisation of pump and probe, and the impact parameter. 
These sets of measurements have been carried out to validate the interferometric amplification of the deflection signal and to characterise the possible sources of systematic measurement bias, particularly the impact of the residual phase noise $\delta \phi(x,y)$. % on the interferometric deflection signal.

\subsection{Parameters in the interaction area}

Measurements presented in this article have been performed with waists at focus in the interaction area $w_0 \simeq W_0\simeq 35$~µm for both the probe and the pump, and may vary from 25 to 40~µm according to the set of measurements. Exact values are specified for each result. 
It corresponds to a Rayleigh length $z_R = \pi w_0^2 / \lambda \simeq 3$~mm. 
The tilt angle is set to $\theta_{\rm tilt} = \ang{5.3}$. This corresponds to an average interaction length $L_{int} \simeq w_0/\theta_{\rm tilt} \simeq 400$~µm, which is $8$ times smaller than the Rayleigh length. This means that the beams are collimated in the interaction zone.
It has been verified that the transverse intensity profiles of the probe and the pump are gaussian along the interaction area with a width in agreement (within 10$\%$) with the expected width in the case of a gaussian beam propagation. 
It has also been verified that the impact parameter is constant along the interaction length. 
Details of the measurements of the transverse intensity profiles at focus are given in Appendix~\ref{ap:focus}. 

For all the measurements, the energy of the probe pulse is equal to $E_{probe}=1$~µJ, corresponding to a peak intensity of the probe in the interaction area equal to $I_{probe} = 1.2\times E_{probe}/(w_0^2 \Delta t) \simeq 1 \ \mathrm{TW/cm}^2$
The intensity of the pump in the interaction area is around $I_{pump}\simeq 5 \ \mathrm{TW/cm}^2$, except for the set of measurements of the deflection signal as a function of the pump intensity where the pump intensity has been increased up to 20~TW/cm$^2$\footnote{The exact energy, waists at focus and intensity of the pump will be specified for each set of measurements in the following.}. 

It has been verified with the numerical simulation (see Appendix~\ref{ap:plasma}) that the contribution of the plasma is negligible for a pump intensity below 20~TW/cm$^2$. 
It has also been verified with the numerical simulation that the contribution of the probe field (i.e. its interference with the pump field) is negligible in the deflection signal, when the pump intensity is above $5 \ \mathrm{TW/cm}^2$.

\subsection{Amplification and sensitivity}

\begin{figure}  
    \centering
    \includegraphics[width=8.5cm]{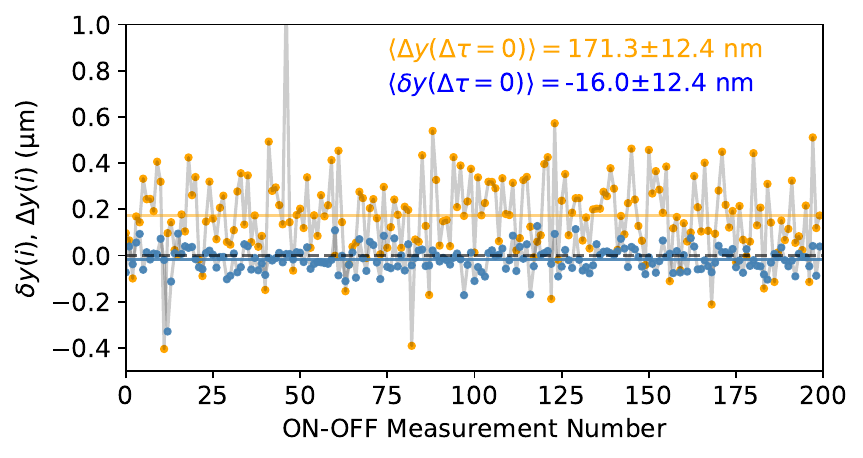}
    \caption{Distributions of 200 successive "ON-OFF" measurements (corresponding to 400 laser shots) of the direct deflection signal $\delta y(i)$ (blue) and the amplified deflection signal $\Delta y(i)$ (orange). Data were taken with a peak intensity of the pump at focus of 4.7~TW/cm$^2$ and with the pump and the probe in time coincidence.}
    \label{fig:data_stat}
\end{figure}

We first present in Figure~\ref{fig:data_stat} an example of the distributions of the ``ON-OFF'' measurement of the direct deflection signal $\delta y(i)$ and the amplified deflection signal $\Delta y(i)$, obtained when the pump and the probe pulses are in time coincidence. The measurements have been performed with a pump energy of $2$ $\mu$J, with pulse durations $\Delta t \simeq 70$ fs, and transverse waists at focus $w_{0,x}=35$~µm and $w_{0,y}=40$~µm for the probe, and $W_{0,x}=24$~µm and $W_{0,y}=30$~µm for the pump corresponding to a peak intensity of the pump at focus of about 5 TW/cm$^2$. The pump and the probe were in time coincidence and the impact parameter $b$ was set to its optimal value $b=b_{opt}$. The pump was vertically shifted below the probe. This corresponds to a negative direct deflection signal and a positive amplified signal (since the amplification factor is negative).
The average measured value of the direct deflection signal is $\langle\delta y\rangle = - 16 \pm 4.75$~nm, while the average amplified signal is $\langle\Delta y\rangle = 171.3 \pm 12.4$~nm.

We observe here a clear amplification of the deflection signal when measuring the interference intensity profile. 
The measured amplification factor is $\mathcal{A} = \langle\Delta y\rangle / \langle\delta y\rangle \simeq -11$. While its sign is negative as expected, its amplitude is lower than the expected value $\mathcal{A_F} = -25$. This difference can be explained by the presence of a non uniform phase noise $\delta \phi(x,y)$, with a non-zero barycenter as explained in the previous section, and discussed in more detail below.

The errors given for the measured values of $\langle\delta y\rangle$ and $\langle\Delta y\rangle$ are purely statistical and equal to $\sigma_y/\sqrt{N}$, where $N$ is the number of ON-OFF measurements, and $\sigma_y$ is the standard deviation of the distributions and corresponds to the spatial resolution.  The measured values are $\sigma_y = 55$~nm for the direct deflection signal and $\sigma_y = 175$~nm for the amplified signal.
On average, the spatial resolution of the amplified measurements carried out in the pilot experiment can vary between $150$ and $300$~nm, while the ultimate shot noise resolution of the present CCD camera is $\sigma_y^{\mathrm{CCD}} = 30$~nm, as measured in~\cite{mailliet2023search_3}. The relatively poor observed resolution for the amplified interferometric signal is due to the phase noise fluctuations induced by the mechanical vibrations of the interferometer~\cite{mailliet2024performance}. Let us note that the interferometer could not be isolated from external vibrations during these measurements.
Details of the current spatial resolution of the interferometer and the description of the method  being developed to measure and suppress the phase noise at high frequency are given in~\cite{mailliet2024performance}. 

The direct deflection signal $\delta y$ has been calculated by a three-dimensional numerical simulation, including the experimental parameters of the present measurements (see Appendix~\ref{ap:plasma}).
The calculated signal is equal to the measured direct deflection signal when the value of the optical Kerr index is set to $n_2 = (1.0 \pm 0.2) 10^{-19}$~cm$^2$/W (only statistical errors). This value is in agreement with the value $n_2 = (1.2 \pm 0.3) 10^{-19}$ cm$^2$/W reported by Loriot \textit{et al.} in~\cite{loriot2009measurement,loriot2010measurement}.

Finally, we may estimate the sensitivity of the current interferometric measurement.
The amplified signal $\langle\Delta y\rangle = 171.3 \pm 12.4$~nm, presented in Figure~\ref{fig:data_stat}, has been measured in 40~seconds of collected data, with a statistical significance of $171.3/12.4 = 13.8$~sigma.
%, i.e. $13.8/\sqrt(40) = 6.3$~sigma per second. 
It has been carried out with a pump intensity $I=5$~TW/cm$^2$ producing an optical index variation $\delta n = n_2 \times I \simeq 5 \times 10^{-7}$. 
This means that the DeLLight pilot experiment can measure an index variation $\delta n = 5 \times 10^{-7}\times\sqrt{40} \ / \ 13.8 = 2.3 \times 10^{-7}$ at 1-sigma confidence level in 1~second of collected data.

\subsection{Time delay}

The direct and amplified deflection signals have been measured as a function of the time delay $\Delta \tau$ between the arrival of the pump and the probe pulses at focus, using the delay stage (DL in Figure~\ref{fig:setup}) of the pump beam.
For this set of measurements, the impact parameter $b$ was set to its optimal value $b=b_{opt}$. 
We have verified that the value of the impact parameter does not change when we vary the time delay of the pump. 
%Measurements have been performed with a pump energy of $2 \ \mu$J and a pulse duration of $70$~fs (FWHM), corresponding to a peak intensity of the pump at focus of about 3~TW/cm$^2$. 
Figure~\ref{fig:results} (a) shows the measured deflection signals $\langle \delta y \rangle$ and $\langle \Delta y \rangle$, as a function of the time delay $\Delta \tau$. 
A positive time delay corresponds to a pump pulse in advance. 
We verify that both the direct and the amplified deflection signals are maximum when the pump and the probe pulses are in time coincidence ($\Delta\tau=0$, this measurement corresponds to the measurement presented in the previous section in Figure~\ref{fig:data_stat}).
We measure an average amplification factor $\mathcal{A} \simeq -11$.
We also verify that the width of the deflection profile is in good agreement with the pulse duration of $70$~fs (FWHM) for both the pump and the probe. 
We also note that the amplitude of the signal decreases when the pump is ahead in time, confirming the absence of plasma induced by the pump.
A dedicated measurement was carried out with the pump pulse delayed by 250~fs, so that the pump and the probe do not interact, in order to verify the absence of signal. The average measured deflection is $\langle \Delta y^{d}\rangle = -7.8 \pm 5.5 $~nm, in agreement with the expected null value.

\subsection{Pump intensity}

\begin{figure}
    \centering
    \includegraphics[width=7.3cm]{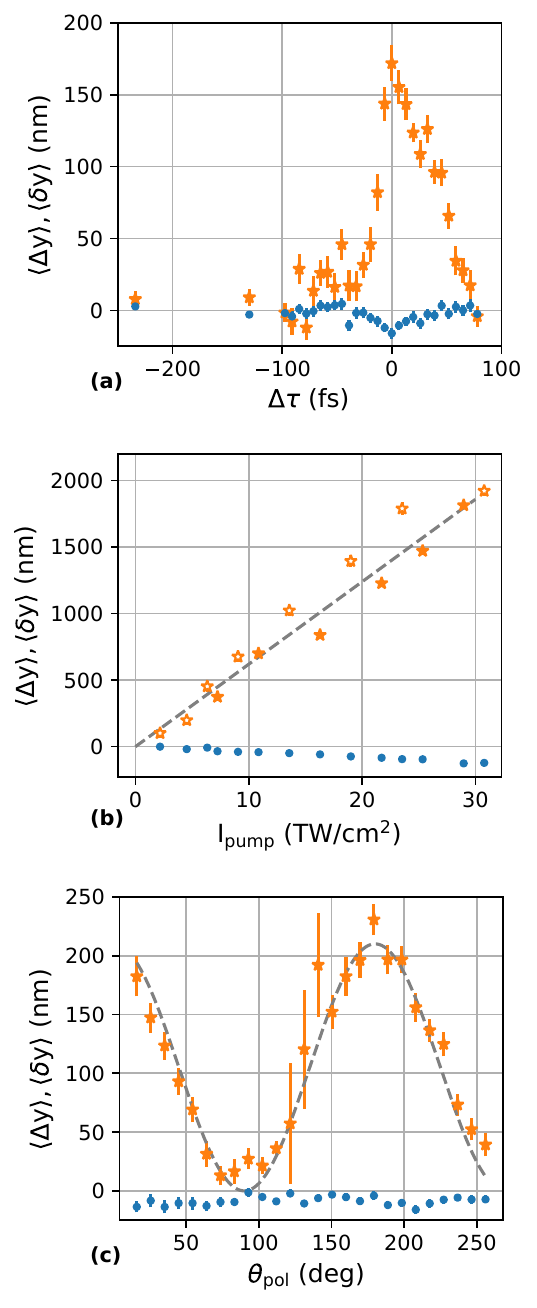}
    \caption{Amplified deflection signal $\Delta y$ (orange stars) and direct deflection signal $\delta y$ (blue dots) measured as a function of (a) the time delay between the probe and the pump pulses, (b) the pump intensity, and (c) the relative polarisation angle $\alpha $ between the pump and probe pulses. Pump and probe pulses have a pulse duration of about $70$ fs (FWHM). (a) Measurements have been performed with a pump energy of $2$ $\mu$J and a waist at focus of the pump $W_{0,x}= 24$~µm, $W_{0,y} = 30$~µm corresponding to a peak intensity of the pump at focus of about 4.7 TW/cm$^2$. (b) Measurements have been performed with a waist at focus of th epump $W_{0,x}= 30$~µm, $W_{0,y} = 33$~µm. (c) Measurements have been performed with a pump energy of $2$ $\mu$J and a waist at focus of the pump $W_{0,x}= 24$~µm, $W_{0,y} = 29$~µm, corresponding to a peak intensity of about 4.9 TW/cm$^2$. }
    \label{fig:results}
\end{figure}

The direct and amplified deflection signals have been measured as a function of the pump intensity at focus in the interaction area. 
The energy of the pump pulse is varied by using a rotating half-wave plate followed by a linear film polarizer (set to the $p$-polarisation of the probe), maintaining the time coincidence of the probe and the pump. 
The measurements have been carried out first by  increasing the pump intensity, then by decreasing it. 
The complete scan took about 90~minutes. 
Results are presented in Figure~\ref{fig:results}(b).
The amplitude of the direct and amplified signals increases linearly with the pump intensity, as expected for a signal induced by first order optical Kerr effect.
However, we measure a small difference of the amplification factor when we compare the results of the first part of the scan when the intensity is being increased ($\mathcal{A} \simeq - 17$, solid orange stars in Figure~\ref{fig:results}) with the results of the second part of the scan when the intensity is being decreased ($\mathcal{A} \simeq -20$, empty orange stars in Figure~\ref{fig:results}). 
This difference is due to the fact that the transverse spatial distribution of the phase noise has slightly drifted during the intensity scan, with the consequence to changed by about 15$\%$ the value of the amplification factor.

%However, we observe that the deflection signal is not exactly proportional to the pump intensity. The reason is that at low pump intensity, where the pump and probe intensities are comparable, the refractive index seen by the probe becomes modulated by the interference between the pump and probe pulses. In that case, the intensity of the probe being constant, the intensity generating the index gradient can be expressed as $I = (E_{\rm pump}+E_{\rm probe})^2 = I_{\rm pump} + a\sqrt{I_{\rm pump}} + b$, where $a = 2\sqrt{I_{\rm probe}}$ and $b= I_{\rm probe}$ are two constants independent of the pump intensity. Since the last term $b$ is identical for the counter-propagating reference pulse, it does not generate any deflection of the interferometric signal in the dark output. Therefore the deflection signal is simply given by:
%\begin{equation}
%    \Delta y = \alpha I_{\rm pump} + \beta\sqrt{I_{\rm pump}}
%\end{equation}
%As shown in Figure~\ref{fig:results}~(b), the fitted function is in good agreement with the data.

%In addition, by comparing the direct and amplified signals, we measure an average amplification factor $\mathcal{A} = \Delta y / \delta y \simeq -18$, about $40\%$ lower than the expected value $\mathcal{A_F} = -25$.

\subsection{Relative polarisation}

The deflection signal $\langle \Delta y\rangle$ was then measured as a function of the relative polarisation angle $\theta_{\rm pol}$ between the pump and the probe.  $\theta_{\rm pol}$ was tuned by rotating the polarisation of the pump pulse with a half-wave plate. 
%Measurements have been performed with a pump energy of $6 \ \mu$J and a pulse duration of about 200~fs, corresponding to a peak intensity of about 3~TW/cm$^2$.
Results are presented in Figure~\ref{fig:results}~(c). 
The measured amplified signal is in good agreement with the fitted function $\Delta y = \Delta y_{\rm max} × \cos^2(\theta_{\rm pol})$, dashed gray line in Figure~\ref{fig:results}.
The amplified signal is maximum when the relative polarisation angle is null, i.e., when the probe and pump polarisations are parallel, and the signal is minimum when their polarisations are orthogonal.  Here, we measure an amplification factor $\mathcal{A} = \Delta y / \delta y \simeq -13$.

\subsection{Impact parameter}

Finally, the direct and amplified deflection signals have been  
measured as a function of the impact parameter $b$, which was varied by rotating vertically the mirror along the pump beam before focusing it (M-3 in Figure \ref{fig:setup}). 
The purpose of this measurement is to analyse more precisely the influence of the phase noise on the measurement of the amplified signal $\Delta y$.

The direct signal $\delta y$ is by definition insensitive to the interference phase noise. It is maximum when $b=b_{opt}$ and null when $b=0$, and varies as~\cite{robertson2021experiment}:
\begin{equation} 
\delta y(b)  = \delta y_{\rm max} \, \frac{b}{b_{\rm opt}} \, e^{ \frac{1}{2} \left( 1 - \left(\frac{b}{b_{\rm opt}}\right)^{2} \right) } 
\label{eq:delta-y-vs-b}
\end{equation}

On the other hand, when the phase noise is not negligible, the amplified deflection signal is not only related to the deflection of the probe but is also sensitive to the phase noise transverse distribution $(\delta \phi(x,y)+\delta\psi/2)^2$ in Equation~(\ref{eqIONgen}), as discussed in section~\ref{sec:amplification} . 
However, the phase delay $\delta \psi$ induced by the pump on the probe is now maximum when $b=0$ and varies as~\cite{robertson2021experiment}:
\begin{equation}
\delta \psi(b)  = \delta \psi_{\rm max} \,  e^{ -\frac{1}{2} \left(\frac{b}{b_{\rm opt}}\right)^{2}  } \,,
\label{eq:delta-psi-vs-b}
\end{equation}
where $\delta \psi_{\rm max}$ is directly related to $\delta y_{\rm max}$~\cite{robertson2021experiment} by:
\begin{equation}
    \delta \psi_{\rm max} = \delta y_{\rm max} \frac{2 \pi b_{\rm opt}}{\lambda f} e^{-\frac{1}{2}} \,,
\label{eq:delta-psi-delta-y}
\end{equation}
where $f$ is the focal length of lenses L1 and L2 and $\lambda$ is the wavelength of the probe pulse.

We can therefore characterize the influence of the phase noise by measuring separately the variations of the direct $\delta y$ and amplified $\Delta y$ deflection signals as a function of the impact parameter $b$.

\begin{figure*}
    \centering
    \includegraphics[width=17cm]{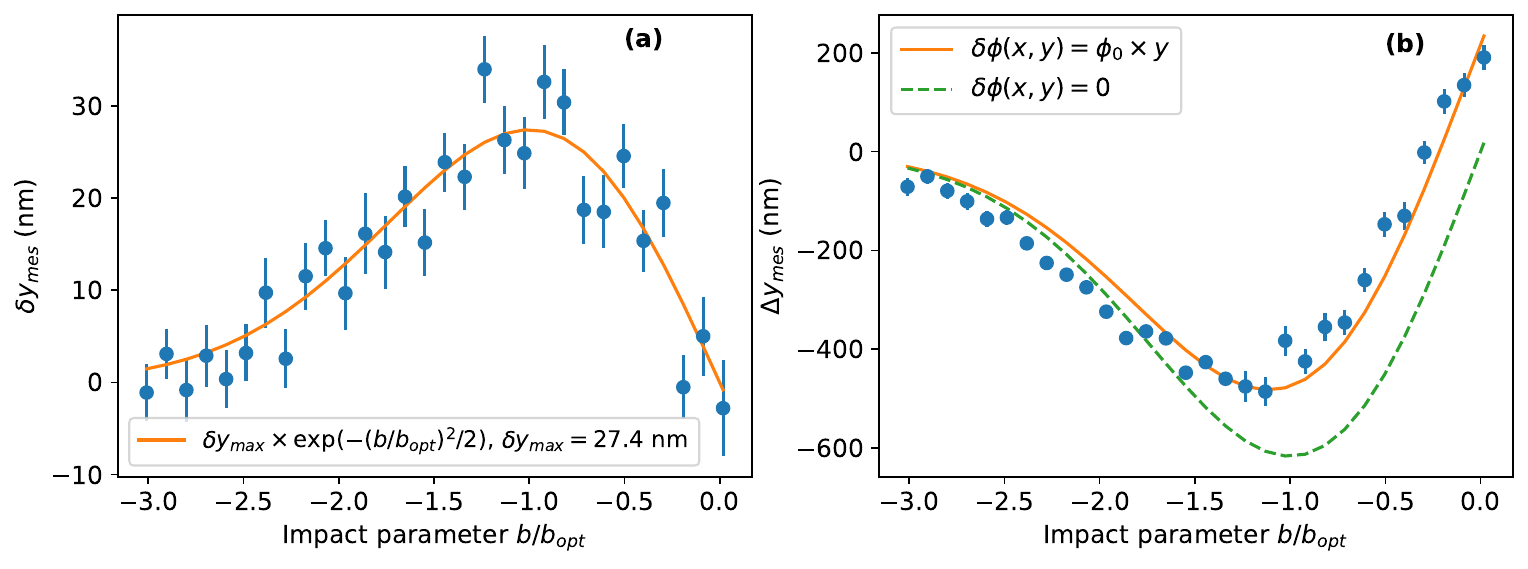}
    \caption{Measurements (in blue dots) of the direct deflection signal $\delta y$ (left panel - (a)) and the amplified deflection signal $\Delta y$ (right panel - (b))  as a function of the normalized impact parameter $b/b_{\mathrm{opt}}$. Measurements have been performed with a pump energy of 6~µJ, a pulse duration of $250$ fs (FWHM), a waist at focus $W_{0,x} = 29$~µm and $W_{0,y} = 33$~µm corresponding to a peak intensity of the pump at focus of about 3 TW/cm$^2$. (a) The solid orange curve is the result of the fit of the function given by Equation~(\ref{eq:delta-y-vs-b}), where $\delta y_{max}$ is the only free parameter.
    (b) The solid orange curve is the result of the fit of the function given by Equations~(\ref{eqIONgen}) and (\ref{eq:delta-psi-vs-b}) where the phase noise $\delta \phi(x,y) = \delta \phi_0  y $ is the only free parameter, and $\delta a$ and $\delta \psi$ are fixed to their measured values with $\delta a =0.02$ and $\delta \psi = 270$~µrad. The fitted value of the phase noise is $\delta \phi_0 = 2.0 \times 10^{-5}$~mrad/nm. The dashed green curve is plotted as a comparison in the absence of phase noise ($\delta \phi_0 = 0 $).
    }
    \label{fig:results2}
\end{figure*}

Figure~\ref{fig:results2}~(a) shows the results of the measurement of the direct deflection signal $\delta y$ as a function of the normalized impact parameter $b/b_{\mathrm{opt}}$. 
The normalisation $b/b_{\rm opt}$ is obtained by fitting the function given in Equation~(\ref{eq:delta-y-vs-b}), where $\delta y_{\rm max}$ is the only free parameter. The result of the fit, shown in Figure~\ref{fig:results2}~(a), is in good agreement with the data and the fitted value is $\delta y_{\rm max} = 27.4$~nm.
From Equation~(\ref{eq:delta-psi-delta-y}) it corresponds to a phase delay signal $\delta \psi_{\rm max} = 270$~µrad.

The measurement of the amplified deflection signal as a function of $b$, shown in Figure~\ref{fig:results2}~(b), is then fitted using Equations~(\ref{eqIONgen}) and (\ref{eq:delta-psi-vs-b}). For this fit, we set $\delta a$ to its measured value $\delta a = 0.02$ and $\delta \psi_{\rm max} = 270$~µrad. 
The phase noise $\delta \phi(x,y)$ is then the only free parameter.
We have assumed for simplicity a phase noise that varies linearly along the vertical axis of the beam: $\delta \phi(x,y) = \delta \phi_0 y $, where $\delta \phi_0$ is the only free parameter of the fit. The result of the fit is in good agreement with the data with a fitted value $\delta \phi_0 = 20$~µrad/µm. 
For comparison, we have also presented the result of the expected function when the phase noise is null ($\delta \phi(x,y) = 0$). 
By comparing the two cases, we see that the presence of a phase noise with a linear transverse distribution reduces the amplitude of the amplified interferometric deflection signal with a reduction factor which increases with $b$. 
Since the direct deflection signal is unaffected by the phase noise, the amplification factor $\mathcal{A}$ can be reduced from its expected value $\mathcal{A}_{F}$. 
This explains why the difference between the expected and measured amplification factors observed on the data presented in Figures~\ref{fig:results}(a) and (c) is attributed to the phase noise. 
It should be noted here that, while the behaviour of the amplified signal is in agreement with the expected result, there is still a residual discrepancy at large impact factor. This can be explained by the technical difficulty of modifying the impact parameter without changing other parameters such as the temporal coherence or the crossing angle between the pump and the probe.

\section{Conclusion}

%The goal of the DeLLight project is to observe for the first time the optical nonlinearity in vacuum, by measuring the refraction of a low-energy focused laser pulse when crossing the refractive index gradient induced by its interaction with a high-energy focused laser pulse. The deflection signal is amplified using a Sagnac interferometric measurement. 

In this article, we have reported the first measurements performed with the DeLLight pilot interferometer of the deflection of light by light in air with a low-energy pump pulse. We have shown that the deflection signal is in agreement with the expected signal induced by the optical Kerr effect in air. Moreover, we have verified that the signal varies as expected as a function of the temporal delay between the pump and the probe, the intensity of the pump, the relative polarisation angle between the probe and the pump, and the impact parameter. 
Furthermore, these measurements validate the DeLLight interferometric amplification method, with an amplification factor $|\mathcal{A}|$ ranging between $10$ to $20$, slightly lower than the expected amplification $|\mathcal{A_F}| = 25$. We have shown that this difference can be explained by the presence of a non-uniform residual phase noise in the interferometer. 
The measured $1\sigma$ sensitivity of the DeLLight pilot experiment is $\delta n \simeq 2.3 \times 10^{-7}$ in 1~second of collected data. This sensitivity is limited by the current amplification factor delivered by the interferometer, and by the spatial resolution $\sigma_y \simeq 200$~nm of the amplified interferometric signal in the dark output of the interferometer. Both the amplification factor and the spatial resolution are in turn limited by the phase noise fluctuations induced by the mechanical vibrations of the interferometer.
%which was not yet isolated from external vibrations.
We are developing a method using a secondary delay probe pulse to monitor and suppress the phase noise at high frequency, in a similar way as the monitoring and suppression of the beam pointing fluctuations.

The DeLLight collaboration is preparing to launch a series of measurements in vacuum with intense pump pulses delivered by the LASERIX facility. Note that this initial series of measurements in vacuum will also provide the opportunity to test the effect of residual gas, as discussed in \cite{robertson2021experiment}. With the energy of 2.5~J for the pump pulse and with a minimum waist of the probe and the pump at focus in the interaction area $w_0 = W_0 = 5 \ \mu$m, the expected variation of the vacuum optical index is $\delta n_{\mathrm{QED}} = 3\times 10^{-13}$~\cite{robertson2021experiment}. With an amplification factor of the interferometer $|\mathcal{A}| = 250$ (corresponding to the best extinction factor achieved with the pilot) and a spatial resolution $\sigma_y = 15$~nm (corresponding to the shot noise of available CCD cameras), the expected QED deflection signal could be observed at a 5~$\sigma$ confidence level with about one month of collected data.
We explain in greater detail in a separate article~\cite{mailliet2024performance} how such amplification and spatial resolution can be achieved. 
From a technological perspective, no show-stoppers has yet been identified to achieving the required sensitivity.

\section{Acknowledgment}
This research is supported by the French National Research Agency through Grant No. ANR-22-CE31-0003-01 for the Advanced-DeLLight project. 
%-------------------------------------------------------
%
%     APPENDIX
%
%-------------------------------------------------------
\appendix

\section{Analytical calculation of the intensity profile of the interference signal in the dark output of the interferometer}
\label{ap:analytical-calculation}

Denoting by $E_0$ the incident field entering the interferometer, the electric fields of the probe and the reference in the dark output of the Sagnac interferometer are defined by:

\begin{equation}
    \left\{ 
        \begin{array}{ll}
            E_{probe}(x,y) &= E_0(x,y - \delta y) e^{-i \delta \psi} (1-\delta a)/2 \\ \\
            E_{ref}(x,y)  &= E_0(x,y) e^{2i \delta \phi} (1+\delta a)/2   e^{-i \pi}
         \end{array}
    \right.
\end{equation}

$\delta y$ is the un-amplified deflection signal of the probe caused by the interaction with the pump. The electric field of the probe includes the effects of the Kerr signal deflection $\delta \phi$ and phase-shift $\delta \psi$ due to the interaction with the pump in the Sagnac interferometer. The probe is reflected two times on the beamsplitter on each side of the beam splitting coating, hence the term in $r^2 = \frac{1}{2}(1-\delta a)$ whereas the reference is transmitted two times (term in $t^2 = \frac{1}{2}(1 + \delta a)$). We arbitrarily choose to put the phase noise term $\delta \phi(x, y)$ in the electric field of the reference. The term $e^{-i\pi}$ in $E_{ref}$ comes from the $\pi$ phase-shift between the probe and the reference.
\begin{widetext}

In order to simplify the calculations, we will note the derivative of the initial electric field $E_0(x,y)$ in amplitude such as: $E'_0(x,y) = \frac{\delta E_0(x,y)}{\delta y}$. 
\begin{equation}
\label{eq:C-1}
    \left\{ 
        \begin{array}{ll}
            2 E_{probe}(x,y) &= (E_0(x,y) - \delta y E'_0(x,y)) (1-\delta a) e^{-i \delta \psi} \\ \\
            2 E_{ref}(x,y)   &= E_0(x,y) e^{2i \delta \phi} (1+\delta a) e^{-i \pi} 
         \end{array}
    \right.
\end{equation}

The intensity profile $I_{out}(x, y)$ of the interference between the probe and the reference in the dark output of the Sagnac interferometer is:
\begin{equation}
        \begin{array}{ll}
            I_{out}(x,y)&= (E_{probe}(x,y) + E_{ref}(x,y)) (E_{probe}(x,y) + E_{ref}(x,y))^{*}

         \end{array}
   % I_{out} = |E_{probe} + E_{ref}|^2 = (E_{probe} + E_{ref}) (E_{probe} + E_{ref})^{*}
\end{equation}
Using Equation~(\ref{eq:C-1}), the intensity profile  $I_{out}(x,y)$  becomes:

\begin{equation}
    \begin{array}{ll}
         4 I_{out}(x,y) &= [(E_0(x,y) - \delta y E'_0(x,y))(1-\delta a)]^2 + [E_0(x,y)(1+\delta a)]^2 \\
         &- 2 \cos(\delta \psi + 2\delta \phi) E_0(x,y)(E_0(x,y) - \delta y E'_0(x,y))(1-(\delta a)^2)
    \end{array}
\end{equation}
%\end{widetext}

Considering the small angles approximation, the cosine term can be simplified such $\cos(\delta \psi + 2\delta \phi) \approx 1 - (\delta \psi + 2\delta \phi)^2/2$. The parameter $\delta a$ correspond to an asymmetry in amplitude between the probe and the reference, while the parameters $\delta \psi$ and $\delta \phi$ correspond to a phase difference. Hence, we separate these two effects and define the amplified intensity profile $I^{ampl}_{out}(x,y)$ in the dark output and the phase term $I^{phase}_{out}(x,y)$, such as:

%\begin{widetext}
\begin{equation}
\label{eq:C-2}
        \begin{array}{ll}
            4 I_{out}(x,y)&= \underbrace{[(E_0(x,y) - \delta y E'_0(x,y))(1-\delta a)]^2 + [E_0(x,y)(1+\delta a)]^2 - 2E_0(x,y)(E_0(x,y) - \delta y E'_0(x,y))(1-(\delta a)^2)}_{I^{ampl}_{out}(x,y)} \\
                            &+ \underbrace{(\delta \psi + 2\delta \phi)^2 E_0(x,y)(E_0(x,y) - \delta y E'_0(x,y))(1-(\delta a)^2)}_{I^{phase}_{out}(x,y)}

         \end{array}
\end{equation}

\end{widetext}

To simplify, we note: $I_{in}(x,y) = E_0(x,y)^2$ ; $I'_{in}(x,y) = \frac{\delta I_{in}(x,y)}{\delta y} = 2E_0(y)E'_0(x,y)$. We develop $I^{ampl}_{out}(x,y)$ and $I^{phase}_{out}(x,y)$ 
(see \cite{mailliet2023search_1} for details) 
to finally obtained:
\begin{equation}
\label{eq:C-3}
        \begin{array}{ll}
                I^{ampl}_{out}(x,y)  &= (\delta a)^2  I_{in}(x,y+ \frac{1-\delta a}{2 \delta a}  \delta y) \\
                \\
                I^{phase}_{out}(x,y) &= (\delta \phi + \delta \psi/2)^2 (1 - (\delta a)^2) I_{in} (x,y - \delta y/2)
         \end{array}
\end{equation}

The final expression of the residual intenisty profile in the dark output of the Sagnac interferometer is obtained by substituting  equation (\ref{eq:C-3}) in (\ref{eq:C-2}) and considering $\delta a \ll 1$. It corresponds to the intensity profile obtained when the probe interacts with the pump ("ON" measurement), so it will be named $I_{ON}(x,y)$:

\begin{align}
\begin{split}
    I_{\rm ON}(x, y) ={}&  (\delta a)^2 I_{\rm in} \left(x,y + \frac{1}{2\delta a} \delta y \right)\\
                    & + \left(\delta \phi(x,y) + \frac{\delta \psi}{2} \right)^2  I_{\rm in}\left(x,y-\frac{\delta y}{2}\right)
\end{split}
\end{align}

When the pump is OFF ("OFF" measurement), the phase shift $\delta \psi$ and the deflection signal $\delta y$ are null ($\delta \psi = \delta y = 0 $) and the "OFF" intensity profile, named $I_{OFF}(x,y)$ is:

\begin{equation}
    I_{\rm OFF}(x, y) = \left( (\delta a)^2 + (\delta \phi(x,y))^2 \right) I_{\rm in}(x,y)
\end{equation}

\section{Description of the main back-reflections on the rear side of the Sagnac beamsplitter}
\label{ap:back-reflections}

\begin{figure}
    \centering
    \includegraphics[width=10cm]{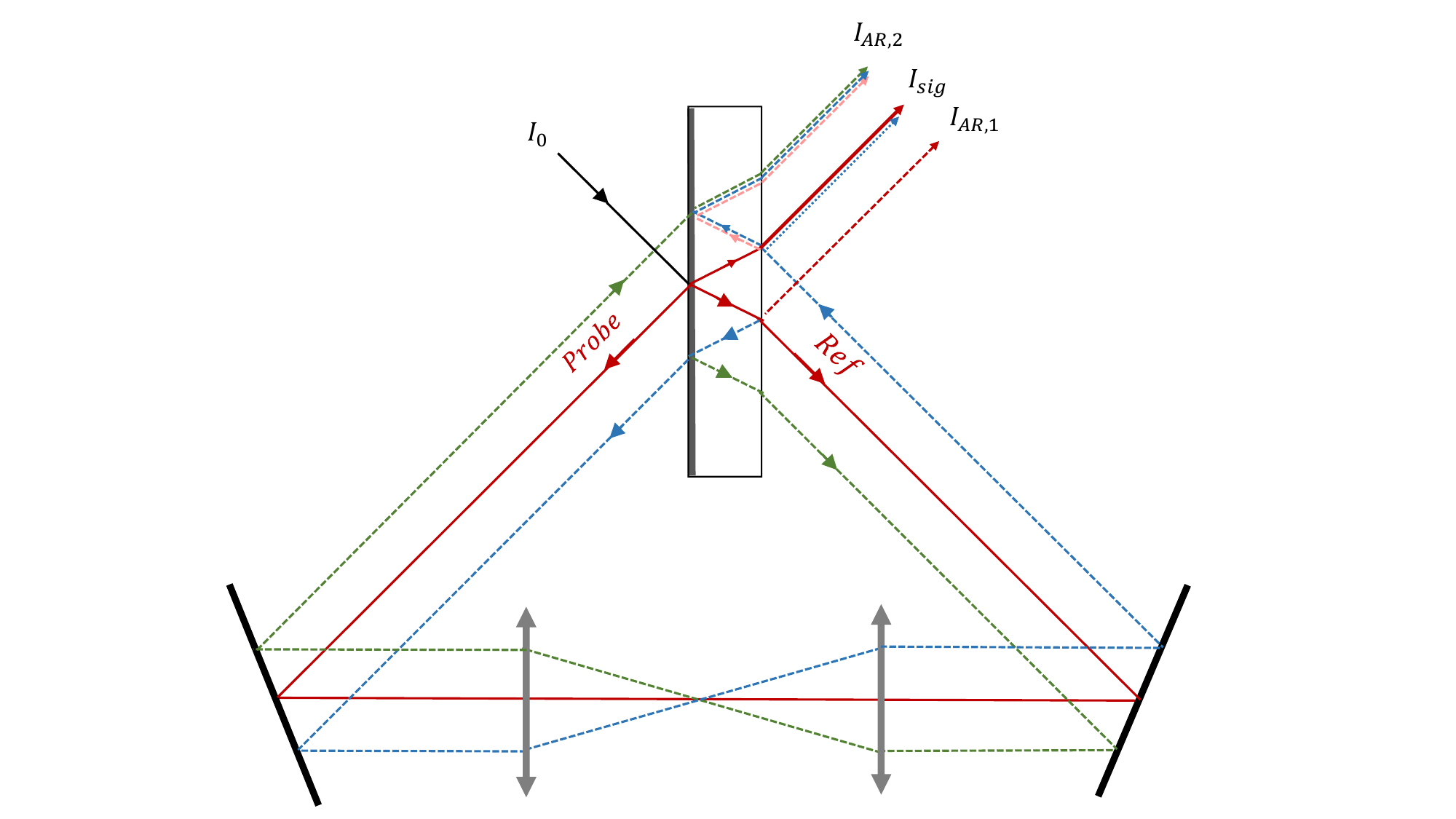}
    \caption{Schematic view of the ray tracing of the Probe and Ref pulses  providing the main interference signal $I_{sig}$ used to measure the amplified deflection signal $\Delta y$ (continuous red line), the direct back-reflection $I_{AR,1}$ used to measure the direct deflection signal $\delta y$ (red dashed line) and the second back-reflection $I_{AR,2}$ used to measure and suppress off-line the beam pointing fluctuations (green and blue dashed line).}
    \label{fig:back-reflexions}
\end{figure}

The ray tracing scheme of the back-reflections on the rear side of the Sagnac beamsplitter is given in Figure~\ref{fig:back-reflexions}.

One back-reflection, named $I_{AR,1}$, corresponds to the direct reflection of the probe pulse on the rear side of the beamsplitter. Since it is the direct image of the  probe beam after circulating in the interferometer, it is also deflected by the pump pulse in the interaction area, but its deflection signal is not amplified via interference with the reference pulse in the dark output. Therefore the measurement of its position on the CCD allows to measure the direct deflection signal $\delta y$ of the probe.
The second back-reflection, named $I_{AR,2}$, results from the constructive interference of two back reflections plus the destructive interference of two other back reflections. 
Their intensities are: 
\begin{equation}
    I_{AR,1}(y) = (r_{AR} \times r)^2 × I_0(y+\delta y)
\end{equation}
\begin{align}
\begin{split}
\label{eq:backside_signal}
    I_{AR,2}(y) ={}& (\underbrace{r_{AR} \times rt^2}_{(1)} + \underbrace{r_{AR} \times rt^2}_{(2)})^2  I_0(y)\\
                 & + \underbrace{(r_{AR} \times r^3 - r_{AR} \times tr^2)}_{(3)}I_0(y)
\end{split}
\end{align}
where $r$ and $t$ are the reflection and transmission coefficients in amplitude of the beamsplitter, $r_{AR}$ is the back-reflection coefficient in amplitude on the rear side of the beamsplitter, and $I_0$ is the incident intensity.
Since $r^2 \simeq t^2 \simeq 1/2$, and if we neglect the phase noise, the intensities of the two back reflections are equal:  $I_{AR,1}= I_{AR,2} = \frac{R_{AR}}{2} \times I_0$

Let us detail the mechanism producing $I_{AR,2}$. When the incident pulse $I_0$ has been transmitted through the entrance face of the beamsplitter, it is reflected on the rear side, it returns on the entrance face of the beamsplitter, and it is then partly transmitted (blue ray in Figure~\ref{fig:back-reflexions}) and partly reflected (green ray in Figure~\ref{fig:back-reflexions}), producing two laterally offset beams (on each side) inside the interferometer (blue and green beams in Figure~\ref{fig:back-reflexions}) 
These two beams interfere constructively when they return to the beam splitter giving the two first terms in Equation~\ref{eq:backside_signal}. The third term is simply produced by the back reflection of the main destructive interference beam $I_{sig}$ followed by the reflection on the entrance face. This third destructive interference term is therefore negligible.
Therefore, the  back-reflection $I_{AR,2}$ corresponds to the constructive interference of the two laterally offset beams (blue and green in Figure~\ref{fig:back-reflexions}) which are not in coincidence with the pump (delayed). This second back-reflection is not affected by the pump and is therefore used as a reference for beam-pointing correction.

\section{Measurements of the transverse intensity profiles at the focus}
\label{ap:focus}

This appendix details the features of the pump and probe pulses at the focus of the interaction area. First will be presented their transverse intensity profiles at focus, and then the longitudinal variation of their waists during the interaction.

\subsection*{Transverse intensity profiles at focus}

To measure the transverse intensity profiles of the pulses, % at focus, 
we insert a mirror between the focusing lens of the probe (L-1) and the interaction area at focus. The mirror reflects the probe and pump pulses before focalisation, which are then collected off-axis onto a high-resolution CCD camera with a small pixel size (Basler acA2500-14gm, pixel dimension: $2.2 \times 2.2 \, \mu {\rm m}^{2}$). The CCD camera is first placed at the focal point of the probe beam where the width of its transverse intensity profile is minimum. This longitudinal position of the CCD camera is labeled as $z = 0$. The longitudinal position of the lens used to focus the pump beam (L-3) is then adjusted in order to minimise the width of the transverse intensity profile of the pump.

\begin{figure}
    \centering
    \includegraphics[width=8cm]{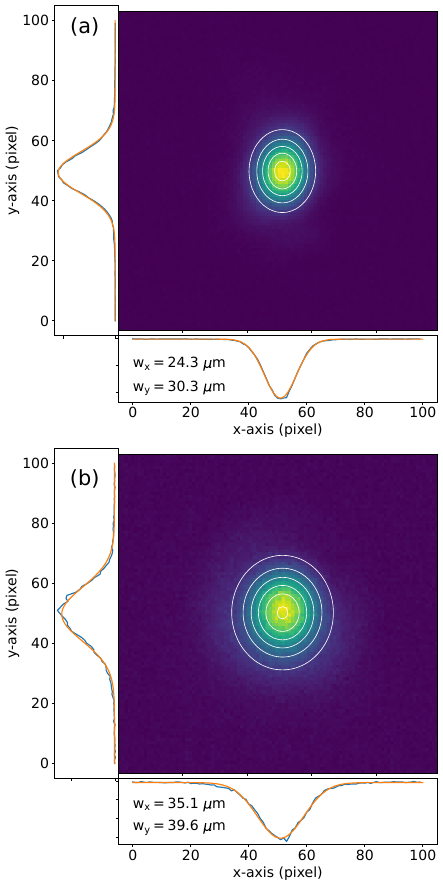}
    \caption{CCD images of the intensity profiles in the interaction area of the pump (a) and the probe (b) in the Sagnac interferometer. A 2-dimensional Gaussian fit delivers the beam waists in x and y-direction}
    \label{fig:spot_at_focus}
\end{figure}

Figure~\ref{fig:spot_at_focus} shows an example of the pump and probe intensity profiles recorded % at focus 
at $z = 0$ after completing the measurements of the deflection signal as a function of the time delay $\Delta \tau$ % between the pulses 
(i.e., those shown in Figure~\ref{fig:results}(a)). For each pulse, a two-dimensional Gaussian profile is fitted to the data to measure the transverse horizontal $w_x$ and vertical $w_y$ waists at focus of both pump and probe. Note that the profiles at focus are well described by two-dimensional Gaussian profiles. In practice, for all measurements performed with the DeLLight pilot experiment, the probe and pump waists at focus range from approximately $25$ to $40$ $\mu$m. % at focus in the interaction area.

\subsection*{Waists as a function of the longitudinal position}

In order to measure the variation of the pump and probe waists as function of the longitudinal position $z$, the CCD camera is longitudinally translated, along the bisector of the two beam, by steps of $\Delta z = 100$~$\mu$m from $-300$~$\mu$m to $400$~$\mu$m, corresponding to a longitudinal scan twice longer than the interaction length $L_{int} \simeq 300$~µm.
For each $z$, we measure the horizontal and vertical waists $w_x$ and $w_y$ of the pump and the probe. 
Results are presented in Figure ~\ref{fig:spot_at_focus}. 
In the same figure, we plot the theoretical waist value $w(z)$ for a gaussian beam propagation, given by $w(z) = w_0 \sqrt{1 + (z/z_R)^2}$ , where $w_0$ is the minimum waist at focus ($z=0$)  and $z_R = \pi w_0^2 / \lambda \simeq 3$~mm is the Rayleigh length. The measured values are in good agreement with the expected one, with a relative difference lower than 5$\%$.
We also verify here that the pump and probe beams are collimated (constant waist) along the interaction length $L_{int} \simeq 300$~µm. 

\begin{figure}
    \centering
    \includegraphics[width=8.6cm]{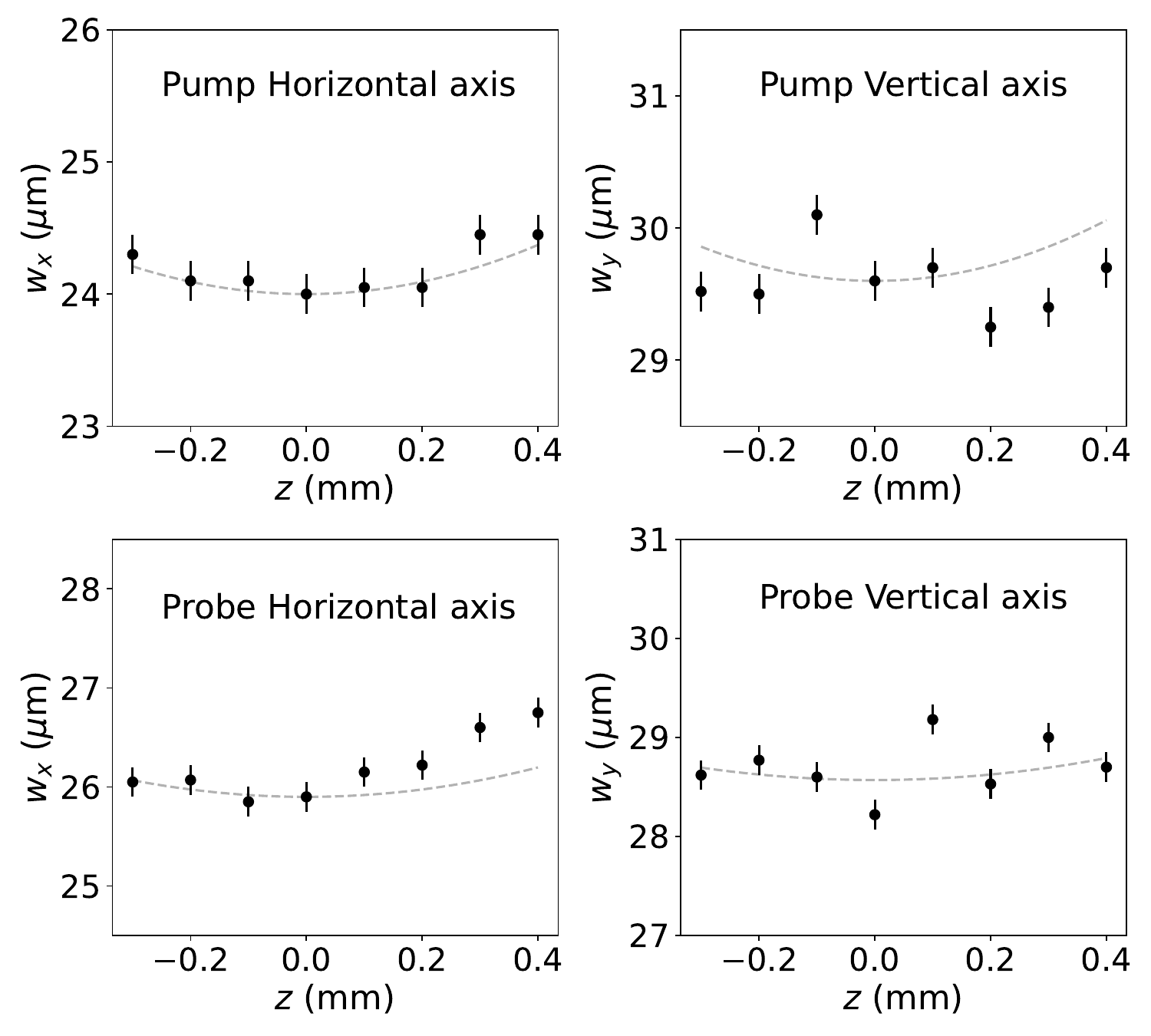}
    \caption{Evolution of the probe (top) and pump (bottom) waists at focus in the $x$- (left) and $y$-directions (right), as a function of the longitudinal position $z$ of the CCD camera (with $z = 0$ the position at focus). The dashed lines correspond to the expected waist at focus for an ideal Gaussian beam.}
    \label{fig:w_f_z}
\end{figure}

\subsection*{Fluctuations of the probe and pump positions at focus}

The pump and probe pulses are delivered (via the beamsplitter (BS-1) on Figure~\ref{fig:setup}) from the same incident beam. Their beam pointing fluctuations are therefore correlated. As a consequence, the relative position fluctuations of the pump and probe intensity profiles at focus must be negligible. In order to test that out, we record a dedicated set of measurements of the pump and probe intensity profiles at focus during $50$ minutes ($3000$ events with an acquisition rate of $1$ Hz).

\begin{figure}
    \centering
    \includegraphics[width=8.6cm]{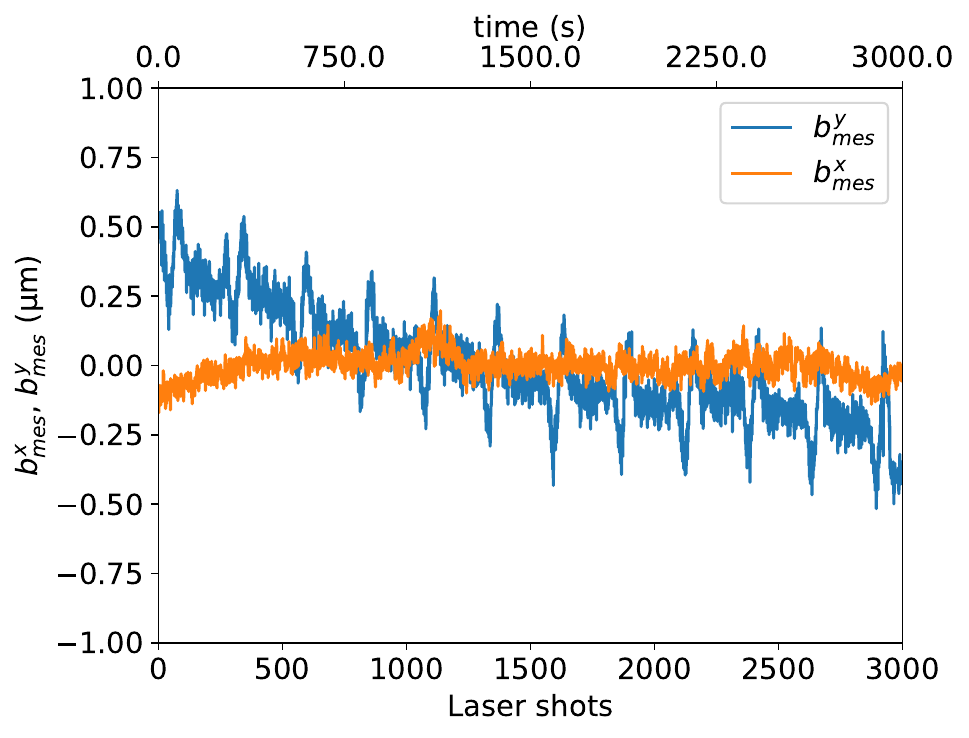}
    \caption{Variation of the horizontal $b^x_{mes}$ and vertical $b^y_{mes}$ impact parameter which corresponds to the relative positions of the probe ($X_g^{Probe}$,$Y_g^{Probe}$) and the pump ($X_g^{Pump}$,$Y_g^{Pump}$) pulses at focus in the interaction area, as a function of time: $b^x_{mes} = X_g^{Pump} - X_g^{Probe}$ and $b^y_{mes} = Y_g^{Pump} - Y_g^{Probe}$. }
    \label{fig:relative_fluctuations_at_focus}
\end{figure}

Figure \ref{fig:relative_fluctuations_at_focus} shows the variation of the relative horizontal and vertical positions of the two beams at focus (i.e. the impact parameter $b^x_{mes}$ and $b^y_{mes}$ respectively) as a function of time. A vertical drift of the relative vertical position between pump and probe of about $750$ nm is measured after 50 minutes of data collection, corresponding to approximately $3$ µm after about $3$ hours, which is typically the duration of a measurement campaign for DeLLight measurements in air. It is one order of magnitude smaller than the waist of the pulses at focus for the pump and the probe, which are around $35$ µm. This drift is therefore negligible.

\section{Numerical simulation code and contribution of the plasma}
\label{ap:plasma}

The DeLLight three-dimensional simulation code solves the Maxwell equations of the propagation of the probe pulse crossing the effective refractive index (optical Kerr index and plasma index) induced in air by the pump pulse.  
The contribution of the probe field can be added by interfering the amplitudes of the probe and pump fields. However, it has been verified that when the intensity of the pump is at least 4 times higher than the probe, then the value of the calculated deflection signal is unmodified if we ignore the contribution of the probe field. Details of the simulation are given in the DeLLight internal note~\cite{NumericalCalculationRobertson}. 

For the optical Kerr effect, we consider only the first order effect, with a Kerr index variation $\delta n_K$ proportional to the intensity of the field: $\delta n_K = n_2 \times I$, with $n_2 = 10^{-19}$~W/cm$^2$.

For the plasma, we use the ionisation rate $r$ of oxygen and nitrogen calculated by Couairon and Mysyrowicz~\cite{couairon2007femtosecond}, computed from the full Keldysh-PPT formulation with a determined pre-factor for diatomic molecules developed by Mishima et al.~\cite{mishima2002generalization}. The generalised Keldysh–PPT formula describes the ionisation rate of a gas in the multi-photon regime (below $\simeq 10^{13}$~W/cm$^2$), and the tunnel regime (above $\simeq 10^{13}$~W/cm$^2$).
The index variation $\delta n_p$ due to the plasma is negative and equal to $\delta n_p = -\sqrt{1-n_c/n_e}$ where $n_c=(\omega^2 m_e)/(\mu_0 c^2 e^2) \simeq 1.7 \times 10^{27}$~m$^{-3}$ is the critical density and $n_e$ is the ionisation electron density. 

\begin{figure}
    \centering
    \includegraphics[width=8cm]{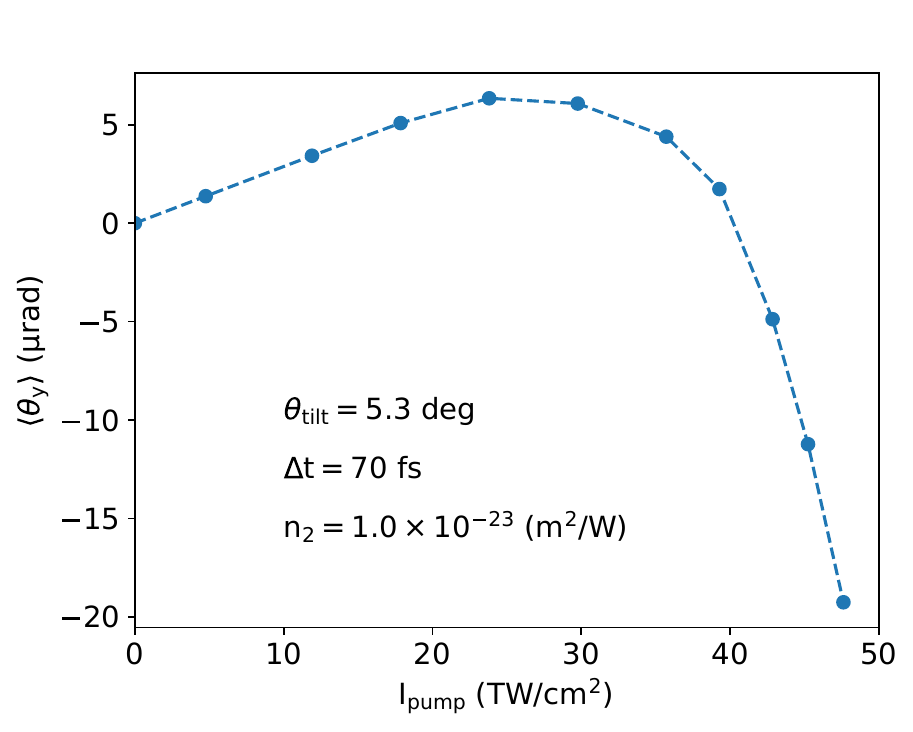}
    \caption{Deflection angle $\langle \theta_y \rangle$ (in µrad) of the probe pulse induced by the pump pulse, as a function of the pump intensity $I_{pump}$ (in TW/cm$^2$), calculated by DeLLight three-dimensional simulation code, including the experimental parameters of the measurements presented in Figure~\ref{fig:data_stat}: $\theta_{tilt}=5.3$~deg, $\Delta t=70$~fs, transverse waists at focus $W_{0,x}=24$~µm and $W_{0,y}=30$~µm for the pump, and $w_{0,x}=35$~µm and $w_{0,y}=40$~µm for the probe.}
    \label{fig:plasma}
\end{figure}

Figure~\ref{fig:plasma} shows the deflection angle $\langle \theta_y \rangle$ of the probe pulse induced by the pump pulse, calculated by the DeLLight three-dimensional simulation code, when we take into account the optical Kerr effect and the plasma. Here, the intensity of the probe is negligible compared with that of the pump. 
We verify that for a pump intensity lower than about 25~TW/cm$^2$, the deflection is only induced by the optical Kerr effect with an expected deflection angle proportional to the pump intensity. Above 25~TW/cm$^2$, the plasma index is not negligible and becomes even the dominant process above 40~TW/cm$^2$ with a negative deflection angle. 

\bibliography{main}

%apsrev4-2.bst 2019-01-14 (MD) hand-edited version of apsrev4-1.bst
%Control: key (0)
%Control: author (8) initials jnrlst
%Control: editor formatted (1) identically to author
%Control: production of article title (0) allowed
%Control: page (0) single
%Control: year (1) truncated
%Control: production of eprint (0) enabled
\providecommand{\noopsort}[1]{}\providecommand{\singleletter}[1]{#1}%
\begin{thebibliography}{41}%
\makeatletter
\providecommand \@ifxundefined [1]{%
 \@ifx{#1\undefined}
}%
\providecommand \@ifnum [1]{%
 \ifnum #1\expandafter \@firstoftwo
 \else \expandafter \@secondoftwo
 \fi
}%
\providecommand \@ifx [1]{%
 \ifx #1\expandafter \@firstoftwo
 \else \expandafter \@secondoftwo
 \fi
}%
\providecommand \natexlab [1]{#1}%
\providecommand \enquote  [1]{``#1''}%
\providecommand \bibnamefont  [1]{#1}%
\providecommand \bibfnamefont [1]{#1}%
\providecommand \citenamefont [1]{#1}%
\providecommand \href@noop [0]{\@secondoftwo}%
\providecommand \href [0]{\begingroup \@sanitize@url \@href}%
\providecommand \@href[1]{\@@startlink{#1}\@@href}%
\providecommand \@@href[1]{\endgroup#1\@@endlink}%
\providecommand \@sanitize@url [0]{\catcode `\\12\catcode `\$12\catcode
  `\&12\catcode `\#12\catcode `\^12\catcode `\_12\catcode `\%12\relax}%
\providecommand \@@startlink[1]{}%
\providecommand \@@endlink[0]{}%
\providecommand \url  [0]{\begingroup\@sanitize@url \@url }%
\providecommand \@url [1]{\endgroup\@href {#1}{\urlprefix }}%
\providecommand \urlprefix  [0]{URL }%
\providecommand \Eprint [0]{\href }%
\providecommand \doibase [0]{https://doi.org/}%
\providecommand \selectlanguage [0]{\@gobble}%
\providecommand \bibinfo  [0]{\@secondoftwo}%
\providecommand \bibfield  [0]{\@secondoftwo}%
\providecommand \translation [1]{[#1]}%
\providecommand \BibitemOpen [0]{}%
\providecommand \bibitemStop [0]{}%
\providecommand \bibitemNoStop [0]{.\EOS\space}%
\providecommand \EOS [0]{\spacefactor3000\relax}%
\providecommand \BibitemShut  [1]{\csname bibitem#1\endcsname}%
\let\auto@bib@innerbib\@empty
%</preamble>
\bibitem [{\citenamefont {Euler}\ and\ \citenamefont
  {Kockel}(1935)}]{euler1935streuung}%
  \BibitemOpen
  \bibfield  {author} {\bibinfo {author} {\bibfnamefont {H.}~\bibnamefont
  {Euler}}\ and\ \bibinfo {author} {\bibfnamefont {B.}~\bibnamefont {Kockel}},\
  }\bibfield  {title} {\bibinfo {title} {{\"U}ber die streuung von licht an
  licht nach der diracschen theorie},\ }\href@noop {} {\bibfield  {journal}
  {\bibinfo  {journal} {Naturwissenschaften}\ }\textbf {\bibinfo {volume}
  {23}},\ \bibinfo {pages} {246} (\bibinfo {year} {1935})}\BibitemShut
  {NoStop}%
\bibitem [{\citenamefont {Heisenberg}\ and\ \citenamefont
  {Euler}(1936)}]{heisenberg1936folgerungen}%
  \BibitemOpen
  \bibfield  {author} {\bibinfo {author} {\bibfnamefont {W.}~\bibnamefont
  {Heisenberg}}\ and\ \bibinfo {author} {\bibfnamefont {H.}~\bibnamefont
  {Euler}},\ }\bibfield  {title} {\bibinfo {title} {Folgerungen aus der
  diracschen theorie des positrons},\ }\href@noop {} {\bibfield  {journal}
  {\bibinfo  {journal} {Zeitschrift f{\"u}r Physik}\ }\textbf {\bibinfo
  {volume} {98}},\ \bibinfo {pages} {714} (\bibinfo {year} {1936})}\BibitemShut
  {NoStop}%
\bibitem [{\citenamefont {Schwinger}(1951)}]{schwinger1951gauge}%
  \BibitemOpen
  \bibfield  {author} {\bibinfo {author} {\bibfnamefont {J.}~\bibnamefont
  {Schwinger}},\ }\bibfield  {title} {\bibinfo {title} {On gauge invariance and
  vacuum polarization},\ }\href@noop {} {\bibfield  {journal} {\bibinfo
  {journal} {Physical Review}\ }\textbf {\bibinfo {volume} {82}},\ \bibinfo
  {pages} {664} (\bibinfo {year} {1951})}\BibitemShut {NoStop}%
\bibitem [{\citenamefont {Burke}\ \emph {et~al.}(1997)\citenamefont {Burke},
  \citenamefont {Field}, \citenamefont {Horton-Smith}, \citenamefont {Spencer},
  \citenamefont {Walz}, \citenamefont {Berridge}, \citenamefont {Bugg},
  \citenamefont {Shmakov}, \citenamefont {Weidemann}, \citenamefont {Bula},\
  and\ \citenamefont {others.}}]{burke1997positron}%
  \BibitemOpen
  \bibfield  {author} {\bibinfo {author} {\bibfnamefont {D.}~\bibnamefont
  {Burke}}, \bibinfo {author} {\bibfnamefont {R.}~\bibnamefont {Field}},
  \bibinfo {author} {\bibfnamefont {G.}~\bibnamefont {Horton-Smith}}, \bibinfo
  {author} {\bibfnamefont {J.}~\bibnamefont {Spencer}}, \bibinfo {author}
  {\bibfnamefont {D.}~\bibnamefont {Walz}}, \bibinfo {author} {\bibfnamefont
  {S.}~\bibnamefont {Berridge}}, \bibinfo {author} {\bibfnamefont
  {W.}~\bibnamefont {Bugg}}, \bibinfo {author} {\bibfnamefont {K.}~\bibnamefont
  {Shmakov}}, \bibinfo {author} {\bibfnamefont {A.}~\bibnamefont {Weidemann}},
  \bibinfo {author} {\bibfnamefont {C.}~\bibnamefont {Bula}},\ and\ \bibinfo
  {author} {\bibnamefont {others.}},\ }\bibfield  {title} {\bibinfo {title}
  {Positron production in multiphoton light-by-light scattering},\ }\href@noop
  {} {\bibfield  {journal} {\bibinfo  {journal} {Physical Review Letters}\
  }\textbf {\bibinfo {volume} {79}},\ \bibinfo {pages} {1626} (\bibinfo {year}
  {1997})}\BibitemShut {NoStop}%
\bibitem [{atl(2017)}]{atlas2017evidence}%
  \BibitemOpen
  \bibfield  {title} {\bibinfo {title} {Evidence for light-by-light scattering
  in heavy-ion collisions with the atlas detector at the lhc},\ }\href@noop {}
  {\bibfield  {journal} {\bibinfo  {journal} {Nature physics}\ }\textbf
  {\bibinfo {volume} {13}},\ \bibinfo {pages} {852} (\bibinfo {year}
  {2017})}\BibitemShut {NoStop}%
\bibitem [{\citenamefont {Sirunyan}\ \emph {et~al.}(2019)\citenamefont
  {Sirunyan}, \citenamefont {Tumasyan}, \citenamefont {Adam}, \citenamefont
  {Ambrogi}, \citenamefont {Asilar}, \citenamefont {Bergauer}, \citenamefont
  {Brandstetter}, \citenamefont {Dragicevic}, \citenamefont {Er{\"o}},
  \citenamefont {Del~Valle},\ and\ \citenamefont
  {others.}}]{sirunyan2019evidence}%
  \BibitemOpen
  \bibfield  {author} {\bibinfo {author} {\bibfnamefont {A.~M.}\ \bibnamefont
  {Sirunyan}}, \bibinfo {author} {\bibfnamefont {A.}~\bibnamefont {Tumasyan}},
  \bibinfo {author} {\bibfnamefont {W.}~\bibnamefont {Adam}}, \bibinfo {author}
  {\bibfnamefont {F.}~\bibnamefont {Ambrogi}}, \bibinfo {author} {\bibfnamefont
  {E.}~\bibnamefont {Asilar}}, \bibinfo {author} {\bibfnamefont
  {T.}~\bibnamefont {Bergauer}}, \bibinfo {author} {\bibfnamefont
  {J.}~\bibnamefont {Brandstetter}}, \bibinfo {author} {\bibfnamefont
  {M.}~\bibnamefont {Dragicevic}}, \bibinfo {author} {\bibfnamefont
  {J.}~\bibnamefont {Er{\"o}}}, \bibinfo {author} {\bibfnamefont {A.~E.}\
  \bibnamefont {Del~Valle}},\ and\ \bibinfo {author} {\bibnamefont {others.}},\
  }\bibfield  {title} {\bibinfo {title} {Evidence for light-by-light scattering
  and searches for axion-like particles in ultraperipheral pbpb collisions at
  snn= 5.02 tev},\ }\href@noop {} {\bibfield  {journal} {\bibinfo  {journal}
  {Physics Letters B}\ }\textbf {\bibinfo {volume} {797}},\ \bibinfo {pages}
  {134826} (\bibinfo {year} {2019})}\BibitemShut {NoStop}%
\bibitem [{\citenamefont {Heinemann}\ \emph {et~al.}(2020)\citenamefont
  {Heinemann}, \citenamefont {Heinzl},\ and\ \citenamefont
  {Ringwald}}]{heinemann2020luxe}%
  \BibitemOpen
  \bibfield  {author} {\bibinfo {author} {\bibfnamefont {B.}~\bibnamefont
  {Heinemann}}, \bibinfo {author} {\bibfnamefont {T.}~\bibnamefont {Heinzl}},\
  and\ \bibinfo {author} {\bibfnamefont {A.}~\bibnamefont {Ringwald}},\
  }\bibfield  {title} {\bibinfo {title} {Luxe: combining high energy and
  intensity to spark the vacuum},\ }\href@noop {} {\bibfield  {journal}
  {\bibinfo  {journal} {Europhysics News}\ }\textbf {\bibinfo {volume} {51}},\
  \bibinfo {pages} {14} (\bibinfo {year} {2020})}\BibitemShut {NoStop}%
\bibitem [{\citenamefont {Abramowicz}\ \emph {et~al.}(2021)\citenamefont
  {Abramowicz}, \citenamefont {Acosta}, \citenamefont {Altarelli},
  \citenamefont {Assmann}, \citenamefont {Bai}, \citenamefont {Behnke},
  \citenamefont {Benhammou}, \citenamefont {Blackburn}, \citenamefont
  {Boogert}, \citenamefont {Borysov},\ and\ \citenamefont
  {others.}}]{abramowicz2021conceptual}%
  \BibitemOpen
  \bibfield  {author} {\bibinfo {author} {\bibfnamefont {H.}~\bibnamefont
  {Abramowicz}}, \bibinfo {author} {\bibfnamefont {U.}~\bibnamefont {Acosta}},
  \bibinfo {author} {\bibfnamefont {M.}~\bibnamefont {Altarelli}}, \bibinfo
  {author} {\bibfnamefont {R.}~\bibnamefont {Assmann}}, \bibinfo {author}
  {\bibfnamefont {Z.}~\bibnamefont {Bai}}, \bibinfo {author} {\bibfnamefont
  {T.}~\bibnamefont {Behnke}}, \bibinfo {author} {\bibfnamefont
  {Y.}~\bibnamefont {Benhammou}}, \bibinfo {author} {\bibfnamefont
  {T.}~\bibnamefont {Blackburn}}, \bibinfo {author} {\bibfnamefont
  {S.}~\bibnamefont {Boogert}}, \bibinfo {author} {\bibfnamefont
  {O.}~\bibnamefont {Borysov}},\ and\ \bibinfo {author} {\bibnamefont
  {others.}},\ }\bibfield  {title} {\bibinfo {title} {Conceptual design report
  for the luxe experiment},\ }\href@noop {} {\bibfield  {journal} {\bibinfo
  {journal} {The European Physical Journal Special Topics}\ }\textbf {\bibinfo
  {volume} {230}},\ \bibinfo {pages} {2445} (\bibinfo {year}
  {2021})}\BibitemShut {NoStop}%
\bibitem [{\citenamefont {Salgado}\ \emph {et~al.}(2021)\citenamefont
  {Salgado}, \citenamefont {Cavanagh}, \citenamefont {Tamburini}, \citenamefont
  {Storey}, \citenamefont {Beyer}, \citenamefont {Bucksbaum}, \citenamefont
  {Chen}, \citenamefont {Di~Piazza}, \citenamefont {Gerstmayr}, \citenamefont
  {Isele},\ and\ \citenamefont {others.}}]{salgado2021single}%
  \BibitemOpen
  \bibfield  {author} {\bibinfo {author} {\bibfnamefont {F.}~\bibnamefont
  {Salgado}}, \bibinfo {author} {\bibfnamefont {N.}~\bibnamefont {Cavanagh}},
  \bibinfo {author} {\bibfnamefont {M.}~\bibnamefont {Tamburini}}, \bibinfo
  {author} {\bibfnamefont {D.}~\bibnamefont {Storey}}, \bibinfo {author}
  {\bibfnamefont {R.}~\bibnamefont {Beyer}}, \bibinfo {author} {\bibfnamefont
  {P.}~\bibnamefont {Bucksbaum}}, \bibinfo {author} {\bibfnamefont
  {Z.}~\bibnamefont {Chen}}, \bibinfo {author} {\bibfnamefont {A.}~\bibnamefont
  {Di~Piazza}}, \bibinfo {author} {\bibfnamefont {E.}~\bibnamefont
  {Gerstmayr}}, \bibinfo {author} {\bibfnamefont {E.}~\bibnamefont {Isele}},\
  and\ \bibinfo {author} {\bibnamefont {others.}},\ }\bibfield  {title}
  {\bibinfo {title} {Single particle detection system for strong-field qed
  experiments},\ }\href@noop {} {\bibfield  {journal} {\bibinfo  {journal} {New
  Journal of Physics}\ }\textbf {\bibinfo {volume} {24}},\ \bibinfo {pages}
  {015002} (\bibinfo {year} {2021})}\BibitemShut {NoStop}%
\bibitem [{\citenamefont {Fedotov}\ \emph {et~al.}(2023)\citenamefont
  {Fedotov}, \citenamefont {Ilderton}, \citenamefont {Karbstein}, \citenamefont
  {King}, \citenamefont {Seipt}, \citenamefont {Taya},\ and\ \citenamefont
  {Torgrimsson}}]{fedotov2023advances}%
  \BibitemOpen
  \bibfield  {author} {\bibinfo {author} {\bibfnamefont {A.}~\bibnamefont
  {Fedotov}}, \bibinfo {author} {\bibfnamefont {A.}~\bibnamefont {Ilderton}},
  \bibinfo {author} {\bibfnamefont {F.}~\bibnamefont {Karbstein}}, \bibinfo
  {author} {\bibfnamefont {B.}~\bibnamefont {King}}, \bibinfo {author}
  {\bibfnamefont {D.}~\bibnamefont {Seipt}}, \bibinfo {author} {\bibfnamefont
  {H.}~\bibnamefont {Taya}},\ and\ \bibinfo {author} {\bibfnamefont
  {G.}~\bibnamefont {Torgrimsson}},\ }\bibfield  {title} {\bibinfo {title}
  {Advances in qed with intense background fields},\ }\href@noop {} {\bibfield
  {journal} {\bibinfo  {journal} {Physics Reports}\ }\textbf {\bibinfo {volume}
  {1010}},\ \bibinfo {pages} {1} (\bibinfo {year} {2023})}\BibitemShut
  {NoStop}%
\bibitem [{\citenamefont {Klein}\ and\ \citenamefont
  {Nigam}(1964)}]{klein1964birefringence}%
  \BibitemOpen
  \bibfield  {author} {\bibinfo {author} {\bibfnamefont {J.~J.}\ \bibnamefont
  {Klein}}\ and\ \bibinfo {author} {\bibfnamefont {B.}~\bibnamefont {Nigam}},\
  }\bibfield  {title} {\bibinfo {title} {Birefringence of the vacuum},\
  }\href@noop {} {\bibfield  {journal} {\bibinfo  {journal} {Physical Review}\
  }\textbf {\bibinfo {volume} {135}},\ \bibinfo {pages} {B1279} (\bibinfo
  {year} {1964})}\BibitemShut {NoStop}%
\bibitem [{\citenamefont {Baier}\ and\ \citenamefont
  {Breitenlohner}(1967)}]{baier1967vacuum}%
  \BibitemOpen
  \bibfield  {author} {\bibinfo {author} {\bibfnamefont {R.}~\bibnamefont
  {Baier}}\ and\ \bibinfo {author} {\bibfnamefont {P.}~\bibnamefont
  {Breitenlohner}},\ }\bibfield  {title} {\bibinfo {title} {The vacuum
  refraction index in the presence of external fields},\ }\href@noop {}
  {\bibfield  {journal} {\bibinfo  {journal} {Il Nuovo Cimento B (1965-1970)}\
  }\textbf {\bibinfo {volume} {47}},\ \bibinfo {pages} {117} (\bibinfo {year}
  {1967})}\BibitemShut {NoStop}%
\bibitem [{\citenamefont {Homma}\ \emph {et~al.}(2011)\citenamefont {Homma},
  \citenamefont {Habs},\ and\ \citenamefont {Tajima}}]{homma2011probing}%
  \BibitemOpen
  \bibfield  {author} {\bibinfo {author} {\bibfnamefont {K.}~\bibnamefont
  {Homma}}, \bibinfo {author} {\bibfnamefont {D.}~\bibnamefont {Habs}},\ and\
  \bibinfo {author} {\bibfnamefont {T.}~\bibnamefont {Tajima}},\ }\bibfield
  {title} {\bibinfo {title} {Probing vacuum birefringence by phase-contrast
  fourier imaging under fields of high-intensity lasers},\ }\href@noop {}
  {\bibfield  {journal} {\bibinfo  {journal} {Applied Physics B}\ }\textbf
  {\bibinfo {volume} {104}},\ \bibinfo {pages} {769} (\bibinfo {year}
  {2011})}\BibitemShut {NoStop}%
\bibitem [{\citenamefont {Heinzl}\ \emph {et~al.}(2006)\citenamefont {Heinzl},
  \citenamefont {Liesfeld}, \citenamefont {Amthor}, \citenamefont {Schwoerer},
  \citenamefont {Sauerbrey},\ and\ \citenamefont
  {Wipf}}]{heinzl2006observation}%
  \BibitemOpen
  \bibfield  {author} {\bibinfo {author} {\bibfnamefont {T.}~\bibnamefont
  {Heinzl}}, \bibinfo {author} {\bibfnamefont {B.}~\bibnamefont {Liesfeld}},
  \bibinfo {author} {\bibfnamefont {K.-U.}\ \bibnamefont {Amthor}}, \bibinfo
  {author} {\bibfnamefont {H.}~\bibnamefont {Schwoerer}}, \bibinfo {author}
  {\bibfnamefont {R.}~\bibnamefont {Sauerbrey}},\ and\ \bibinfo {author}
  {\bibfnamefont {A.}~\bibnamefont {Wipf}},\ }\bibfield  {title} {\bibinfo
  {title} {On the observation of vacuum birefringence},\ }\href@noop {}
  {\bibfield  {journal} {\bibinfo  {journal} {Optics communications}\ }\textbf
  {\bibinfo {volume} {267}},\ \bibinfo {pages} {318} (\bibinfo {year}
  {2006})}\BibitemShut {NoStop}%
\bibitem [{\citenamefont {Di~Piazza}\ \emph {et~al.}(2005)\citenamefont
  {Di~Piazza}, \citenamefont {Hatsagortsyan},\ and\ \citenamefont
  {Keitel}}]{di2005harmonic}%
  \BibitemOpen
  \bibfield  {author} {\bibinfo {author} {\bibfnamefont {A.}~\bibnamefont
  {Di~Piazza}}, \bibinfo {author} {\bibfnamefont {K.~Z.}\ \bibnamefont
  {Hatsagortsyan}},\ and\ \bibinfo {author} {\bibfnamefont {C.~H.}\
  \bibnamefont {Keitel}},\ }\bibfield  {title} {\bibinfo {title} {Harmonic
  generation from laser-driven vacuum},\ }\href@noop {} {\bibfield  {journal}
  {\bibinfo  {journal} {Physical Review D}\ }\textbf {\bibinfo {volume} {72}},\
  \bibinfo {pages} {085005} (\bibinfo {year} {2005})}\BibitemShut {NoStop}%
\bibitem [{\citenamefont {Lundin}\ \emph {et~al.}(2006)\citenamefont {Lundin},
  \citenamefont {Marklund}, \citenamefont {Lundstr{\"o}m}, \citenamefont
  {Brodin}, \citenamefont {Collier}, \citenamefont {Bingham}, \citenamefont
  {Mendon{\c{c}}a},\ and\ \citenamefont {Norreys}}]{lundin2006analysis}%
  \BibitemOpen
  \bibfield  {author} {\bibinfo {author} {\bibfnamefont {J.}~\bibnamefont
  {Lundin}}, \bibinfo {author} {\bibfnamefont {M.}~\bibnamefont {Marklund}},
  \bibinfo {author} {\bibfnamefont {E.}~\bibnamefont {Lundstr{\"o}m}}, \bibinfo
  {author} {\bibfnamefont {G.}~\bibnamefont {Brodin}}, \bibinfo {author}
  {\bibfnamefont {J.}~\bibnamefont {Collier}}, \bibinfo {author} {\bibfnamefont
  {R.}~\bibnamefont {Bingham}}, \bibinfo {author} {\bibfnamefont
  {J.}~\bibnamefont {Mendon{\c{c}}a}},\ and\ \bibinfo {author} {\bibfnamefont
  {P.}~\bibnamefont {Norreys}},\ }\bibfield  {title} {\bibinfo {title}
  {Analysis of four-wave mixing of high-power lasers for the detection of
  elastic photon-photon scattering},\ }\href@noop {} {\bibfield  {journal}
  {\bibinfo  {journal} {Physical Review A}\ }\textbf {\bibinfo {volume} {74}},\
  \bibinfo {pages} {043821} (\bibinfo {year} {2006})}\BibitemShut {NoStop}%
\bibitem [{\citenamefont {King}\ \emph {et~al.}(2010)\citenamefont {King},
  \citenamefont {Di~Piazza},\ and\ \citenamefont {Keitel}}]{king2010double}%
  \BibitemOpen
  \bibfield  {author} {\bibinfo {author} {\bibfnamefont {B.}~\bibnamefont
  {King}}, \bibinfo {author} {\bibfnamefont {A.}~\bibnamefont {Di~Piazza}},\
  and\ \bibinfo {author} {\bibfnamefont {C.~H.}\ \bibnamefont {Keitel}},\
  }\bibfield  {title} {\bibinfo {title} {Double-slit vacuum polarization
  effects in ultraintense laser fields},\ }\href@noop {} {\bibfield  {journal}
  {\bibinfo  {journal} {Physical Review A}\ }\textbf {\bibinfo {volume} {82}},\
  \bibinfo {pages} {032114} (\bibinfo {year} {2010})}\BibitemShut {NoStop}%
\bibitem [{\citenamefont {Tommasini}\ and\ \citenamefont
  {Michinel}(2010)}]{tommasini2010light}%
  \BibitemOpen
  \bibfield  {author} {\bibinfo {author} {\bibfnamefont {D.}~\bibnamefont
  {Tommasini}}\ and\ \bibinfo {author} {\bibfnamefont {H.}~\bibnamefont
  {Michinel}},\ }\bibfield  {title} {\bibinfo {title} {Light by light
  diffraction in vacuum},\ }\href@noop {} {\bibfield  {journal} {\bibinfo
  {journal} {Physical Review A}\ }\textbf {\bibinfo {volume} {82}},\ \bibinfo
  {pages} {011803} (\bibinfo {year} {2010})}\BibitemShut {NoStop}%
\bibitem [{\citenamefont {Di~Piazza}\ \emph {et~al.}(2012)\citenamefont
  {Di~Piazza}, \citenamefont {M{\"u}ller}, \citenamefont {Hatsagortsyan},\ and\
  \citenamefont {Keitel}}]{di2012extremely}%
  \BibitemOpen
  \bibfield  {author} {\bibinfo {author} {\bibfnamefont {A.}~\bibnamefont
  {Di~Piazza}}, \bibinfo {author} {\bibfnamefont {C.}~\bibnamefont
  {M{\"u}ller}}, \bibinfo {author} {\bibfnamefont {K.~Z.}\ \bibnamefont
  {Hatsagortsyan}},\ and\ \bibinfo {author} {\bibfnamefont {C.~H.}\
  \bibnamefont {Keitel}},\ }\bibfield  {title} {\bibinfo {title} {Extremely
  high-intensity laser interactions with fundamental quantum systems},\
  }\href@noop {} {\bibfield  {journal} {\bibinfo  {journal} {Reviews of Modern
  Physics}\ }\textbf {\bibinfo {volume} {84}},\ \bibinfo {pages} {1177}
  (\bibinfo {year} {2012})}\BibitemShut {NoStop}%
\bibitem [{\citenamefont {Narozhny}\ and\ \citenamefont
  {Fedotov}(2015)}]{narozhny2015extreme}%
  \BibitemOpen
  \bibfield  {author} {\bibinfo {author} {\bibfnamefont {N.}~\bibnamefont
  {Narozhny}}\ and\ \bibinfo {author} {\bibfnamefont {A.}~\bibnamefont
  {Fedotov}},\ }\bibfield  {title} {\bibinfo {title} {Extreme light physics},\
  }\href@noop {} {\bibfield  {journal} {\bibinfo  {journal} {Contemporary
  Physics}\ }\textbf {\bibinfo {volume} {56}},\ \bibinfo {pages} {249}
  (\bibinfo {year} {2015})}\BibitemShut {NoStop}%
\bibitem [{\citenamefont {King}\ and\ \citenamefont
  {Heinzl}(2016)}]{king2016measuring}%
  \BibitemOpen
  \bibfield  {author} {\bibinfo {author} {\bibfnamefont {B.}~\bibnamefont
  {King}}\ and\ \bibinfo {author} {\bibfnamefont {T.}~\bibnamefont {Heinzl}},\
  }\bibfield  {title} {\bibinfo {title} {Measuring vacuum polarization with
  high-power lasers},\ }\href@noop {} {\bibfield  {journal} {\bibinfo
  {journal} {High Power Laser Science and Engineering}\ }\textbf {\bibinfo
  {volume} {4}},\ \bibinfo {pages} {e5} (\bibinfo {year} {2016})}\BibitemShut
  {NoStop}%
\bibitem [{\citenamefont {Ejlli}\ \emph {et~al.}(2020)\citenamefont {Ejlli},
  \citenamefont {Della~Valle}, \citenamefont {Gastaldi}, \citenamefont
  {Messineo}, \citenamefont {Pengo}, \citenamefont {Ruoso},\ and\ \citenamefont
  {Zavattini}}]{ejlli2020pvlas}%
  \BibitemOpen
  \bibfield  {author} {\bibinfo {author} {\bibfnamefont {A.}~\bibnamefont
  {Ejlli}}, \bibinfo {author} {\bibfnamefont {F.}~\bibnamefont {Della~Valle}},
  \bibinfo {author} {\bibfnamefont {U.}~\bibnamefont {Gastaldi}}, \bibinfo
  {author} {\bibfnamefont {G.}~\bibnamefont {Messineo}}, \bibinfo {author}
  {\bibfnamefont {R.}~\bibnamefont {Pengo}}, \bibinfo {author} {\bibfnamefont
  {G.}~\bibnamefont {Ruoso}},\ and\ \bibinfo {author} {\bibfnamefont
  {G.}~\bibnamefont {Zavattini}},\ }\href@noop {} {\bibfield  {journal}
  {\bibinfo  {journal} {Physics Reports}\ }\textbf {\bibinfo {volume} {871}},\
  \bibinfo {pages} {1} (\bibinfo {year} {2020})}\BibitemShut {NoStop}%
\bibitem [{\citenamefont {Cad{\`e}ne}\ \emph {et~al.}(2014)\citenamefont
  {Cad{\`e}ne}, \citenamefont {Berceau}, \citenamefont {Fouch{\'e}},
  \citenamefont {Battesti},\ and\ \citenamefont {Rizzo}}]{cadene2014vacuum}%
  \BibitemOpen
  \bibfield  {author} {\bibinfo {author} {\bibfnamefont {A.}~\bibnamefont
  {Cad{\`e}ne}}, \bibinfo {author} {\bibfnamefont {P.}~\bibnamefont {Berceau}},
  \bibinfo {author} {\bibfnamefont {M.}~\bibnamefont {Fouch{\'e}}}, \bibinfo
  {author} {\bibfnamefont {R.}~\bibnamefont {Battesti}},\ and\ \bibinfo
  {author} {\bibfnamefont {C.}~\bibnamefont {Rizzo}},\ }\bibfield  {title}
  {\bibinfo {title} {Vacuum magnetic linear birefringence using pulsed fields:
  status of the bmv experiment},\ }\href@noop {} {\bibfield  {journal}
  {\bibinfo  {journal} {The European Physical Journal D}\ }\textbf {\bibinfo
  {volume} {68}},\ \bibinfo {pages} {1} (\bibinfo {year} {2014})}\BibitemShut
  {NoStop}%
\bibitem [{\citenamefont {Fan}\ \emph {et~al.}(2017)\citenamefont {Fan},
  \citenamefont {Kamioka}, \citenamefont {Inada}, \citenamefont {Yamazaki},
  \citenamefont {Namba}, \citenamefont {Asai}, \citenamefont {Omachi},
  \citenamefont {Yoshioka}, \citenamefont {Kuwata-Gonokami}, \citenamefont
  {Matsuo},\ and\ \citenamefont {others.}}]{fan2017oval}%
  \BibitemOpen
  \bibfield  {author} {\bibinfo {author} {\bibfnamefont {X.}~\bibnamefont
  {Fan}}, \bibinfo {author} {\bibfnamefont {S.}~\bibnamefont {Kamioka}},
  \bibinfo {author} {\bibfnamefont {T.}~\bibnamefont {Inada}}, \bibinfo
  {author} {\bibfnamefont {T.}~\bibnamefont {Yamazaki}}, \bibinfo {author}
  {\bibfnamefont {T.}~\bibnamefont {Namba}}, \bibinfo {author} {\bibfnamefont
  {S.}~\bibnamefont {Asai}}, \bibinfo {author} {\bibfnamefont {J.}~\bibnamefont
  {Omachi}}, \bibinfo {author} {\bibfnamefont {K.}~\bibnamefont {Yoshioka}},
  \bibinfo {author} {\bibfnamefont {M.}~\bibnamefont {Kuwata-Gonokami}},
  \bibinfo {author} {\bibfnamefont {A.}~\bibnamefont {Matsuo}},\ and\ \bibinfo
  {author} {\bibnamefont {others.}},\ }\href@noop {} {\bibfield  {journal}
  {\bibinfo  {journal} {The European Physical Journal D}\ }\textbf {\bibinfo
  {volume} {71}},\ \bibinfo {pages} {1} (\bibinfo {year} {2017})}\BibitemShut
  {NoStop}%
\bibitem [{\citenamefont {Chen}\ \emph {et~al.}(2007)\citenamefont {Chen},
  \citenamefont {Mei},\ and\ \citenamefont {Ni}}]{chen2007q}%
  \BibitemOpen
  \bibfield  {author} {\bibinfo {author} {\bibfnamefont {S.-J.}\ \bibnamefont
  {Chen}}, \bibinfo {author} {\bibfnamefont {H.-H.}\ \bibnamefont {Mei}},\ and\
  \bibinfo {author} {\bibfnamefont {W.-T.}\ \bibnamefont {Ni}},\ }\bibfield
  {title} {\bibinfo {title} {Q \& a experiment to search for vacuum dichroism,
  pseudoscalar--photon interaction and millicharged fermions},\ }\href@noop {}
  {\bibfield  {journal} {\bibinfo  {journal} {Modern Physics Letters A}\
  }\textbf {\bibinfo {volume} {22}},\ \bibinfo {pages} {2815} (\bibinfo {year}
  {2007})}\BibitemShut {NoStop}%
\bibitem [{\citenamefont {Boyd}(2008)}]{boyd2008nonlinear}%
  \BibitemOpen
  \bibfield  {author} {\bibinfo {author} {\bibfnamefont {R.~W.}\ \bibnamefont
  {Boyd}},\ }\href@noop {} {\emph {\bibinfo {title} {Nonlinear Optics, Third
  Edition}}},\ \bibinfo {edition} {3rd}\ ed.\ (\bibinfo  {publisher} {Academic
  Press, Inc.},\ \bibinfo {address} {USA},\ \bibinfo {year} {2008})\BibitemShut
  {NoStop}%
\bibitem [{\citenamefont {Sarazin}\ \emph {et~al.}(2016)\citenamefont
  {Sarazin}, \citenamefont {Couchot}, \citenamefont {Djannati-Ata{\"\i}},
  \citenamefont {Guilbaud}, \citenamefont {Kazamias}, \citenamefont {Pittman},\
  and\ \citenamefont {Urban}}]{sarazin2016refraction}%
  \BibitemOpen
  \bibfield  {author} {\bibinfo {author} {\bibfnamefont {X.}~\bibnamefont
  {Sarazin}}, \bibinfo {author} {\bibfnamefont {F.}~\bibnamefont {Couchot}},
  \bibinfo {author} {\bibfnamefont {A.}~\bibnamefont {Djannati-Ata{\"\i}}},
  \bibinfo {author} {\bibfnamefont {O.}~\bibnamefont {Guilbaud}}, \bibinfo
  {author} {\bibfnamefont {S.}~\bibnamefont {Kazamias}}, \bibinfo {author}
  {\bibfnamefont {M.}~\bibnamefont {Pittman}},\ and\ \bibinfo {author}
  {\bibfnamefont {M.}~\bibnamefont {Urban}},\ }\bibfield  {title} {\bibinfo
  {title} {Refraction of light by light in vacuum},\ }\href@noop {} {\bibfield
  {journal} {\bibinfo  {journal} {The European Physical Journal D}\ }\textbf
  {\bibinfo {volume} {70}},\ \bibinfo {pages} {1} (\bibinfo {year}
  {2016})}\BibitemShut {NoStop}%
\bibitem [{\citenamefont {Robertson}\ \emph {et~al.}(2021)\citenamefont
  {Robertson}, \citenamefont {Mailliet}, \citenamefont {Sarazin}, \citenamefont
  {Couchot}, \citenamefont {Baynard}, \citenamefont {Demailly}, \citenamefont
  {Pittman}, \citenamefont {Djannati-Ata{\"\i}}, \citenamefont {Kazamias},\
  and\ \citenamefont {Urban}}]{robertson2021experiment}%
  \BibitemOpen
  \bibfield  {author} {\bibinfo {author} {\bibfnamefont {S.}~\bibnamefont
  {Robertson}}, \bibinfo {author} {\bibfnamefont {A.}~\bibnamefont {Mailliet}},
  \bibinfo {author} {\bibfnamefont {X.}~\bibnamefont {Sarazin}}, \bibinfo
  {author} {\bibfnamefont {F.}~\bibnamefont {Couchot}}, \bibinfo {author}
  {\bibfnamefont {E.}~\bibnamefont {Baynard}}, \bibinfo {author} {\bibfnamefont
  {J.}~\bibnamefont {Demailly}}, \bibinfo {author} {\bibfnamefont
  {M.}~\bibnamefont {Pittman}}, \bibinfo {author} {\bibfnamefont
  {A.}~\bibnamefont {Djannati-Ata{\"\i}}}, \bibinfo {author} {\bibfnamefont
  {S.}~\bibnamefont {Kazamias}},\ and\ \bibinfo {author} {\bibfnamefont
  {M.}~\bibnamefont {Urban}},\ }\bibfield  {title} {\bibinfo {title}
  {Experiment to observe an optically induced change of the vacuum index},\
  }\href@noop {} {\bibfield  {journal} {\bibinfo  {journal} {Physical Review
  A}\ }\textbf {\bibinfo {volume} {103}},\ \bibinfo {pages} {023524} (\bibinfo
  {year} {2021})}\BibitemShut {NoStop}%
\bibitem [{\citenamefont {Aharonov}\ \emph {et~al.}(1988)\citenamefont
  {Aharonov}, \citenamefont {Albert},\ and\ \citenamefont
  {Vaidman}}]{aharonov1988result}%
  \BibitemOpen
  \bibfield  {author} {\bibinfo {author} {\bibfnamefont {Y.}~\bibnamefont
  {Aharonov}}, \bibinfo {author} {\bibfnamefont {D.~Z.}\ \bibnamefont
  {Albert}},\ and\ \bibinfo {author} {\bibfnamefont {L.}~\bibnamefont
  {Vaidman}},\ }\bibfield  {title} {\bibinfo {title} {How the result of a
  measurement of a component of the spin of a spin-1/2 particle can turn out to
  be 100},\ }\href@noop {} {\bibfield  {journal} {\bibinfo  {journal} {Physical
  review letters}\ }\textbf {\bibinfo {volume} {60}},\ \bibinfo {pages} {1351}
  (\bibinfo {year} {1988})}\BibitemShut {NoStop}%
\bibitem [{\citenamefont {Aharonov}\ \emph {et~al.}(2010)\citenamefont
  {Aharonov}, \citenamefont {Popescu},\ and\ \citenamefont
  {Tollaksen}}]{aharonov2010time}%
  \BibitemOpen
  \bibfield  {author} {\bibinfo {author} {\bibfnamefont {Y.}~\bibnamefont
  {Aharonov}}, \bibinfo {author} {\bibfnamefont {S.}~\bibnamefont {Popescu}},\
  and\ \bibinfo {author} {\bibfnamefont {J.}~\bibnamefont {Tollaksen}},\
  }\bibfield  {title} {\bibinfo {title} {A time-symmetric formulation of
  quantum mechanics},\ }\href@noop {} {\bibfield  {journal} {\bibinfo
  {journal} {Physics today}\ }\textbf {\bibinfo {volume} {63}},\ \bibinfo
  {pages} {27} (\bibinfo {year} {2010})}\BibitemShut {NoStop}%
\bibitem [{\citenamefont {Dixon}\ \emph {et~al.}(2009)\citenamefont {Dixon},
  \citenamefont {Starling}, \citenamefont {Jordan},\ and\ \citenamefont
  {Howell}}]{dixon2009ultrasensitive}%
  \BibitemOpen
  \bibfield  {author} {\bibinfo {author} {\bibfnamefont {P.~B.}\ \bibnamefont
  {Dixon}}, \bibinfo {author} {\bibfnamefont {D.~J.}\ \bibnamefont {Starling}},
  \bibinfo {author} {\bibfnamefont {A.~N.}\ \bibnamefont {Jordan}},\ and\
  \bibinfo {author} {\bibfnamefont {J.~C.}\ \bibnamefont {Howell}},\ }\bibfield
   {title} {\bibinfo {title} {Ultrasensitive beam deflection measurement via
  interferometric weak value amplification},\ }\href@noop {} {\bibfield
  {journal} {\bibinfo  {journal} {Physical review letters}\ }\textbf {\bibinfo
  {volume} {102}},\ \bibinfo {pages} {173601} (\bibinfo {year}
  {2009})}\BibitemShut {NoStop}%
\bibitem [{\citenamefont {Egan}\ and\ \citenamefont
  {Stone}(2012)}]{egan2012weak}%
  \BibitemOpen
  \bibfield  {author} {\bibinfo {author} {\bibfnamefont {P.}~\bibnamefont
  {Egan}}\ and\ \bibinfo {author} {\bibfnamefont {J.~A.}\ \bibnamefont
  {Stone}},\ }\bibfield  {title} {\bibinfo {title} {Weak-value thermostat with
  0.2 mk precision},\ }\href@noop {} {\bibfield  {journal} {\bibinfo  {journal}
  {Optics letters}\ }\textbf {\bibinfo {volume} {37}},\ \bibinfo {pages} {4991}
  (\bibinfo {year} {2012})}\BibitemShut {NoStop}%
\bibitem [{\citenamefont {Robertson}(2019)}]{robertson2019optical}%
  \BibitemOpen
  \bibfield  {author} {\bibinfo {author} {\bibfnamefont {S.}~\bibnamefont
  {Robertson}},\ }\bibfield  {title} {\bibinfo {title} {Optical kerr effect in
  vacuum},\ }\href@noop {} {\bibfield  {journal} {\bibinfo  {journal} {Physical
  Review A}\ }\textbf {\bibinfo {volume} {100}},\ \bibinfo {pages} {063831}
  (\bibinfo {year} {2019})}\BibitemShut {NoStop}%
\bibitem [{\citenamefont
  {Mailliet}(2023{\natexlab{a}})}]{mailliet2023search_3}%
  \BibitemOpen
  \bibfield  {author} {\bibinfo {author} {\bibfnamefont {A.~M.}\ \bibnamefont
  {Mailliet}},\ }\href@noop {} {Ph.D. thesis},\ \bibinfo  {school}
  {Universit{\'e} Paris-Saclay} (\bibinfo {year} {2023}{\natexlab{a}}),\
  \bibinfo {note} {p. 75-89}\BibitemShut {NoStop}%
\bibitem [{\citenamefont {Mailliet}\ \emph {et~al.}(2024)\citenamefont
  {Mailliet}, \citenamefont {Kraych}, \citenamefont {Couchot}, \citenamefont
  {Sarazin}, \citenamefont {Baynard}, \citenamefont {Demailly}, \citenamefont
  {Pittman}, \citenamefont {Djannati-Ataï}, \citenamefont {Kazamias},
  \citenamefont {Robertson},\ and\ \citenamefont
  {Urban}}]{mailliet2024performance}%
  \BibitemOpen
  \bibfield  {author} {\bibinfo {author} {\bibfnamefont {A.~M.}\ \bibnamefont
  {Mailliet}}, \bibinfo {author} {\bibfnamefont {A.~E.}\ \bibnamefont
  {Kraych}}, \bibinfo {author} {\bibfnamefont {F.}~\bibnamefont {Couchot}},
  \bibinfo {author} {\bibfnamefont {X.}~\bibnamefont {Sarazin}}, \bibinfo
  {author} {\bibfnamefont {E.}~\bibnamefont {Baynard}}, \bibinfo {author}
  {\bibfnamefont {J.}~\bibnamefont {Demailly}}, \bibinfo {author}
  {\bibfnamefont {M.}~\bibnamefont {Pittman}}, \bibinfo {author} {\bibfnamefont
  {A.}~\bibnamefont {Djannati-Ataï}}, \bibinfo {author} {\bibfnamefont
  {S.}~\bibnamefont {Kazamias}}, \bibinfo {author} {\bibfnamefont
  {S.}~\bibnamefont {Robertson}},\ and\ \bibinfo {author} {\bibfnamefont
  {M.}~\bibnamefont {Urban}},\ }\href@noop {} {\bibinfo {title} {Performance of
  a sagnac interferometer to observe vacuum optical nonlinearity}} (\bibinfo
  {year} {2024}),\ \Eprint {https://arxiv.org/abs/2401.13720} {arXiv:2401.13720
  [physics.optics]} \BibitemShut {NoStop}%
\bibitem [{\citenamefont {Loriot}\ \emph {et~al.}(2009)\citenamefont {Loriot},
  \citenamefont {Hertz}, \citenamefont {Faucher},\ and\ \citenamefont
  {Lavorel}}]{loriot2009measurement}%
  \BibitemOpen
  \bibfield  {author} {\bibinfo {author} {\bibfnamefont {V.}~\bibnamefont
  {Loriot}}, \bibinfo {author} {\bibfnamefont {E.}~\bibnamefont {Hertz}},
  \bibinfo {author} {\bibfnamefont {O.}~\bibnamefont {Faucher}},\ and\ \bibinfo
  {author} {\bibfnamefont {B.}~\bibnamefont {Lavorel}},\ }\bibfield  {title}
  {\bibinfo {title} {Measurement of high order kerr refractive index of major
  air components},\ }\href@noop {} {\bibfield  {journal} {\bibinfo  {journal}
  {Optics express}\ }\textbf {\bibinfo {volume} {17}},\ \bibinfo {pages}
  {13429} (\bibinfo {year} {2009})}\BibitemShut {NoStop}%
\bibitem [{\citenamefont {Loriot}\ \emph {et~al.}(2010)\citenamefont {Loriot},
  \citenamefont {Hertz}, \citenamefont {Faucher},\ and\ \citenamefont
  {Lavorel}}]{loriot2010measurement}%
  \BibitemOpen
  \bibfield  {author} {\bibinfo {author} {\bibfnamefont {V.}~\bibnamefont
  {Loriot}}, \bibinfo {author} {\bibfnamefont {E.}~\bibnamefont {Hertz}},
  \bibinfo {author} {\bibfnamefont {O.}~\bibnamefont {Faucher}},\ and\ \bibinfo
  {author} {\bibfnamefont {B.}~\bibnamefont {Lavorel}},\ }\bibfield  {title}
  {\bibinfo {title} {Measurement of high order kerr refractive index of major
  air components: erratum},\ }\href@noop {} {\bibfield  {journal} {\bibinfo
  {journal} {Optics Express}\ }\textbf {\bibinfo {volume} {18}},\ \bibinfo
  {pages} {3011} (\bibinfo {year} {2010})}\BibitemShut {NoStop}%
\bibitem [{\citenamefont
  {Mailliet}(2023{\natexlab{b}})}]{mailliet2023search_1}%
  \BibitemOpen
  \bibfield  {author} {\bibinfo {author} {\bibfnamefont {A.~M.}\ \bibnamefont
  {Mailliet}},\ }\href@noop {} {Ph.D. thesis},\ \bibinfo  {school}
  {Universit{\'e} Paris-Saclay} (\bibinfo {year} {2023}{\natexlab{b}}),\
  \bibinfo {note} {p. 24-30}\BibitemShut {NoStop}%
\bibitem [{\citenamefont {Robertson}()}]{NumericalCalculationRobertson}%
  \BibitemOpen
  \bibfield  {author} {\bibinfo {author} {\bibfnamefont {S.}~\bibnamefont
  {Robertson}},\ }\bibfield  {title} {\bibinfo {title} {Dellight internal note:
  Testing the dellight three-dimensional simulation code},\ }\href@noop {}
  {\bibinfo  {journal}
  {https://groups.ijclab.in2p3.fr/projetdellight/publications}\ }\BibitemShut
  {NoStop}%
\bibitem [{\citenamefont {Couairon}\ and\ \citenamefont
  {Mysyrowicz}(2007)}]{couairon2007femtosecond}%
  \BibitemOpen
\bibfield  {journal} {  }\bibfield  {author} {\bibinfo {author} {\bibfnamefont
  {A.}~\bibnamefont {Couairon}}\ and\ \bibinfo {author} {\bibfnamefont
  {A.}~\bibnamefont {Mysyrowicz}},\ }\bibfield  {title} {\bibinfo {title}
  {Femtosecond filamentation in transparent media},\ }\href@noop {} {\bibfield
  {journal} {\bibinfo  {journal} {Physics reports}\ }\textbf {\bibinfo {volume}
  {441}},\ \bibinfo {pages} {47} (\bibinfo {year} {2007})}\BibitemShut
  {NoStop}%
\bibitem [{\citenamefont {Mishima}\ \emph {et~al.}(2002)\citenamefont
  {Mishima}, \citenamefont {Hayashi}, \citenamefont {Yi}, \citenamefont {Lin},
  \citenamefont {Selzle},\ and\ \citenamefont
  {Schlag}}]{mishima2002generalization}%
  \BibitemOpen
  \bibfield  {author} {\bibinfo {author} {\bibfnamefont {K.}~\bibnamefont
  {Mishima}}, \bibinfo {author} {\bibfnamefont {M.}~\bibnamefont {Hayashi}},
  \bibinfo {author} {\bibfnamefont {J.}~\bibnamefont {Yi}}, \bibinfo {author}
  {\bibfnamefont {S.}~\bibnamefont {Lin}}, \bibinfo {author} {\bibfnamefont
  {H.}~\bibnamefont {Selzle}},\ and\ \bibinfo {author} {\bibfnamefont
  {E.}~\bibnamefont {Schlag}},\ }\bibfield  {title} {\bibinfo {title}
  {Generalization of keldysh’s theory},\ }\href@noop {} {\bibfield  {journal}
  {\bibinfo  {journal} {Physical Review A}\ }\textbf {\bibinfo {volume} {66}},\
  \bibinfo {pages} {033401} (\bibinfo {year} {2002})}\BibitemShut {NoStop}%
\end{thebibliography}%

\end{document}